# Modeling TCP Throughput with Random Packet Drops

Daniel Zaragoza[*] – November 2013

**Abstract.**

*The present report deals with the modeling of the long-term throughput, a.k.a., send rate, of the Transmission Control Protocol (TCP) under the following assumptions. (i) We consider a single 'infinite source' using a network path from sender to receiver. (ii) Each TCP packet is randomly dropped with probability $p$; independently of previous drops or any other event/parameter. (iii) The – never changing – receiver window limits the amount of outstanding data. (iv) The receiver acknowledges every packet. (v) The TCP modeled here conforms to the publicly available standards (RFCs) as concerns congestion control. We validate and determine the limits of the different models proposed here using packet-level simulations. The contributions of the present work are the following: (a) We determine three regimes, and their conditions of applicability, depending on $p$: Linear law regime, square root law regime, and timeout regime. (b) As concerns the relationship between the linear and square root regimes, we give additional insights relatively to previously published work. (c) We give the exact equations governing the TCP send rate in any regime. (d) From the exact equation and under the further condition that the path is not saturated, we give and discuss approximations for the send rate of the NewReno variant of TCP. A by-product of these calculations is the distribution of the sender window, independently of any timing or saturation consideration. (e) These approximations give results that are accurate to a few percent when compared to simulation results. Detailed comparison and sources of errors between theory and simulations are also discussed.*

***Contents***.

# 1. Introduction

The present report deals with the modeling of the long-term throughput, a.k.a., send rate, of the Transmission Control Protocol (TCP) under the following assumptions. (i) We consider a single 'infinite source' using a network path from sender to receiver. (ii) Each TCP packet is randomly dropped with probability $p$; independently of previous drops or any other event/parameter. (iii) The – never changing – receiver window limits the amount of outstanding data. (iv) The receiver acknowledges every packet. (v) The TCP modeled here conforms to the publicly available standards (RFCs) as concerns congestion control; the (many) other aspects of the TCP protocol are left aside. We validate and determine the limits of the different models proposed here using packet-level simulations.

---
[*] daniel.zaragoza@laposte.net

The contributions of the present work are:

(i) We determine three regimes, and their conditions of applicability, depending on $p$: Linear law regime, square root law regime, and timeout regime. Except for the timeout regime, expressions are simple enough to allow back-of-an-envelope calculations. Details are given in section 4.

(ii) As concerns the relationship between the linear and square root regimes, we give additional insights relatively to our previously published work. See section 4 also.

(iii) We give the exact equation governing the TCP send rate in any regime. The model requires the calculation of 16 quantities. These quantities depend on the TCP version considered and some can be simple to obtain. Details are given in section 5.

(iv) From the exact equation and under the further condition that the path is not saturated (definition in section 3.2), we give and discuss approximations for the send rate of the NewReno variant of TCP. A by-product of these calculations is the distribution of the sender window, independently of any timing or saturation consideration. See section 5 also.

(v) These approximations give results that are accurate to a few percent when compared to simulation results. Detailed comparison and sources of errors between theory and simulations are discussed in section 6.

Besides the above mentioned sections, in the next section we discuss related modeling work. In section 3, we give definitions. In particular, in section 3.1, we provide TCP background information and in section 3.2, we give the network-related definitions and the settings used for the simulations. The conclusion and discussion are given in section 7.

In the remainder of the report, we use the following abbreviations and notations.

> Acknowledgement(s) is abbreviated as ack(s).
> Duplicate acknowledgement is abbreviated as DA (we do not use delayed acknowledgements).
> Triple duplicate acknowledgement is abbreviated as TD. A TD triggers a fast retransmit.
> A successful fast recovery is abbreviated as FR.
> Timeout is abbreviated as TO.
> Congestion avoidance is abbreviated as CA.
> Round trip time is abbreviated as RTT.
> Independent and identically distributed is abbreviated as i.i.d.
> We use the words throughput and send rate as synonymous.

> $W_{CA}$, window at which the sender switches from slow-start to congestion avoidance.
> $W_{FR}$, sender window just after FR occurred and "normal" operation resumes in CA.
> $W_R$, receiver window, a.k.a. $rwnd$.
> $S$, slow start threshold, a.k.a. $ssthresh$.
> $\theta$, amount of data outstanding at the sender.
> $W$, sender window.
> $\sigma = W - \theta$, amount to send.
> $RTT$, round trip time. For a given network setting, $RTT$ is a constant independent of possible queuing created by the connection.
> $\lfloor x \rfloor$, is the integer part of $x$.

## 2. Related work

We discuss here related work on the modeling of long TCP connections.

Mathis *et al.* [1], proposed and discussed the applicability of the – famous – square root law. To quote the paper: "*The model applies when TCP performance is determined solely by the congestion avoidance algorithm.*" In the paper, $p$ is the ratio of the number of congestion avoidance events to the number of acknowledged packets. RTT is the average value as measured at the sender and includes possible queuing delays. The receiver window is never a limiting factor. Here, $p$ is the – externally given – packet drop probability and RTT is a constant independent of possible queuing created by the connection. We extend to the lower end the domain of applicability of the law by giving a $p_{min}$. The high end, where the timeout regime starts, is the same for $p$ above 1%. We take into account the receiver window and give a linear law (in $p$) together with a limit $p_{max}$ below which the linear law is applicable. We also study the case where the link (path) is saturated with a limited $W_R$ but packet drops are never due to buffer overflow. The result of our study is the determination of three regimes as discussed in section 4.3 and illustrated in Figure 5.

Padhye *et al.* [2] extend the square root model by taking timeouts into account for the Reno variant of TCP. The model is known as the PFTK model. They provide analytic expressions, which are also used in TCP-friendly protocols. They compare the results of their model to recorded and analyzed transfers performed on the Internet with an implementation using the Reno variant. They define loss indications as TDs and TOs. The ratio of the number of loss indications over the number of packet sent is taken as an approximation of $p$. They also discuss the different assumption used in their model. Not included in the model are the subtleties of FR and the time spent in slow-start. In their model, the path is never saturated. For $p$ sufficiently small their model recovers the square root law. The packet loss process is simplified so that if one packet is lost in a 'round' (a RTT) the following packets in the round are also lost. The first loss in a round (with probability $p$) is i.i.d. from round to round. The RTT is independent of $W$. The technique used in the PFTK model is that of Markov regenerative process with rewards. In less technical terms, the send rate is obtained by calculating the average window during a cycle divided by the average duration of a cycle. A cycle includes TD and TOs events. The PFTK model is revisited by Dunaytsev *et al.* in [3] to include fast retransmit and the slow-start after a timeout.

Altman *et al.* [4] and Barakat [5] use a different mathematical technique to obtain the send rate for long connection. The model accounts for "*any correlation and any distribution of inter-loss times.*" When timeouts have no or little influence, the square root law is improved to account for the variance and correlation function of the inter-loss event process. To quote the paper, "*the classical square root formula is generalized to the case of stationary ergodic losses.*" In that paper also, $p$ is the ratio of the number of loss events to the number packets sent (equation (8) in [3]). Both papers use the results of [1] to account for timeouts. Both papers emphasize the importance of separate modeling of the TCP window evolution (dependence between RTT and window, window limitation, fluid vs. discrete models) and the modeling of the network (diversity of loss event processes). Similarly to [1], model and assumptions are validated against analyzed traces of transfers over the Internet with an actual implementation of NewReno.

A number of other mathematical techniques have been used for modeling the throughput of TCP. An example of technique is Markov chains [6].

Using the same technique as [1], Parvez *et al.* [7] model the send rate of TCP NewReno. They model the fast recovery of NewReno, the timeout behavior, and they use a two-parameter $(p, q)$ loss model to capture the frequency and burstiness of the loss process. $p$ is the packet drop probability and $q$ is the probability of drop of the following packets in the same RTT as the first drop ($q = 1$ in the PFTK model). Their model is analytic and gives an expression for the send rate of NewReno in their equation (29). Surprisingly, the equation shows a $1/p$, rather than a $1/\sqrt{p}$, term in the numerator, which indicates a faster growth to infinity with decreasing $p$. They validate the model using various simulation scenarios and "*All the losses in a single window of data are counted as one loss event. The loss event rate p is taken to be the ratio of the total number of loss events to the total number of segment transmissions, in the period of interest. The average round-trip time R was measured at the sender, and RTO was approximated as 4R.*"

From a high point of view, in the present work, we are interested in obtaining insights, through modeling, on the performance obtainable with the flow and congestion control algorithms used in TCP. For that purpose, we let the receiver window, which governs flow control, be fixed throughout the connection. We also let the receiver acknowledge every packet instead of using delayed acks. The delayed acks strategy is known to degrade the send rate; besides the fact that the relationship RTT/delay timer introduce complications for a correct modeling. As concerns the congestion control part, which effect is to reduce the send rate when a loss is detected, we use the simplest – and easiest to implement – drop policy; that is, i.i.d. random packet drops. When a packet appears, it is dropped with probability $p$ and passes with probability $q = 1 - p$, independently of previous events or any parameter. In previous work, we have used a less simple renewal error process in which errors occur according to a renewal process independently of whether a packet is present or not. We have not seen much difference compared to the simple random drop policy. Additionally, following [4] and [5], performance in CA is better when the loss process – seen by the sender – has high variance and/or shows correlation. As a final general remark, the parameters $p$ (the loss event rate) and $RTT$ used in the literature are those measured at the sender. Here, our point of view is that $p$ (the drop probability or drop rate) and $RTT$ are externally given.

Turning to details, the full model approach developed in section 5 is exact. Under the further assumption that the connection does not saturate the path we develop approximations for the NewReno variant of TCP, which follows the AIMD (additive increase multiplicative decrease) congestion control principle. That assumption may be relaxed under conditions discussed in section 6. In the present approach, the operation of TCP is decomposed in 5 basic elements. For a given TCP version, the determination of the send rate requires the calculation of 16 quantities, a number of which are simple to obtain. Our approach does not require to take into account the slow-start phase after a timeout. It does not require sophisticated stochastic process machinery (at least, in the approximations we give); however, it requires a microscopic (packet-level) analysis of TCP operation and subtleties.

To conclude this section, we note that the number of TCP variants in operation on the Internet has greatly increased between the measurements performed by Medina *et al.* [8] in 2005 and those performed by Yang *et al.* [9] in 2011 (their figure 1 presents a nice synthetic overview of TCP congestion controls). From Rewaskar *et al.* [10] we learn that implementations in the popular operating systems Linux and Windows have a $minRTO$ of 200 ms instead of the 1 second specified in the standard (see next section). The 'vagaries' of real world transfers using TCP are analyzed by Qian *et al.* [11]. Finally, Mathis [12] (as of 2009) note that as ISPs (Internet Service Providers) exert more control on the traffic traversing their networks a paradigm shift (from TCP-friendly to something else) is needed. To quote his words: "*Needless to say our time tested model* (the square root law) *will no longer generally apply. C'est la vie*". Overall, the papers mentioned here tend to indicate that the job of modeling the performance of TCP will be a tough one.

## 3. Definitions

### 3.1 TCP background

In this section we develop the aspects of TCP that are the most relevant to our purpose. (i) TCP is a window-based transport protocol. (ii) It provides flow-control. (iii) It provides congestion control. (iv) Retransmission timeouts are important for its performance.

RFC4614 [13] provides a roadmap to TCP-related documents, as of 2006. Documents of interest here that update previous ones cited in RFC4614 are: RFC5681 [14], – TCP congestion control –, RFC6582 [15] – TCP NewReno –, and RFC6298 [16] – TCP RTO calculation.

TCP is first specified in RFC793 [17]; further details are clarified in RFC1122 [18].

TCP is a window-based protocol that provides a reliable, byte-oriented, data transfer between processes which, most commonly, run on different computers (hosts) themselves located on different networks. In the original specification, TCP can send and receive variable-length "segments". However, most implementations refrain from sending variable length segments unless some conditions are met, e.g., sending a segment empties the send buffer or the Nagle's algorithm is disabled [19].

TCP provides reliability "*by assigning a sequence number to each octet transmitted, and requiring a positive acknowledgment (ACK) from the receiving TCP. If the ACK is not received within a timeout interval, the data is retransmitted.*" Further, "*The sequence number of the first octet of data in a segment is transmitted with that segment and is called the segment sequence number. Segments also carry an acknowledgment number which is the sequence number of the next expected data octet of transmissions in the reverse direction.*" How the timeout is calculated is specified in RFC6298 and is discussed below. In the present work, we use a TCP that is packet-oriented; that is the basic unit of transmission is the packet rather that the octet.

TCP is a window-based protocol which provides flow control to avoid a sender overwhelming a receiver. For that purpose, "*The receiving TCP reports a "window" to the sending TCP. This window specifies the number of octets, starting with the acknowledgment number, that the receiving TCP is currently prepared to receive.*" At the sender and at any time, the amount of data outstanding and not yet acknowledged – a.k.a. the "flight size" – is $\theta = nxt - una$, where $nxt$, is the sequence number of the next unit to send and $una$ is the sequence number of the still unacknowledged unit. Under no circumstances and at no time $\theta > W_R$, where $W_R$ is the receiver window, a.k.a. $rwnd$. $W_R$ is not necessarily fixed and varies according to the speed at which the application "reads" the received data during the course of a connection. Note that $una$ changes only with advancing acks while $nxt$ increases when sending new data and decreases back to $una$ when retransmitting missing data.

A form of indirect/implicit flow control may also exist at the sender. RFC793 states: "*When the TCP transmits a segment containing data, it puts a copy on a retransmission queue and starts a timer; when the acknowledgment for that data is received, the segment is deleted from the queue. If the acknowledgment is not received before the timer runs out, the segment is retransmitted.*" In practice, implementations [19], maintain a "send buffer" (a.k.a. the socket buffer) which contains the unacknowledged data and which is fed with data from the user buffer. The size of this buffer may have impact on the TCP performance. In the present work we do not deal with this matter and assume ideal operation at the sender.

TCP also provides congestion control (RFC5681) by way of the slow-start and congestion avoidance algorithms. For that purpose two new variables are used: the congestion window, $cwnd$ and the slow-start threshold, $ssthresh$, noted $S$ here. A TCP switches between slow-start and congestion avoidance according to the relative values of $cwnd$ and $S$. More precisely, "*The slow start algorithm is used when cwnd < ssthresh, while the congestion avoidance algorithm is used when cwnd > ssthresh. When cwnd and ssthresh are equal, the sender may use either slow start or congestion avoidance.*" Further, "*The initial value of ssthresh SHOULD be set arbitrarily high (e.g.,to the size of the largest possible advertised window), but ssthresh MUST be reduced in response to congestion.*" Here, we use congestion avoidance when $cwnd > S$ meaning that $W_{CA} = S + 1$. Rephrasing RFC5681, during slow-start $cwnd$ is increased by one for each increasing ack received; during congestion avoidance $cwnd$ is increased by one for each $cwnd$ of increasing acks received. To improve performance when the receiver acknowledges every other segment and to discourage misbehaving receivers, a TCP may use Appropriate Byte Counting as per RFC3465 [20]. In our TCP, the receiver acknowledges every segment received and the sender counts the increasing arriving acks before increasing the congestion window in congestion avoidance; this is to avoid numerical rounding errors.

The sending of data (new or retransmission) by TCP is governed by the following equations.

$$W = \min(cwnd, W_R), \theta = nxt - una, \text{ and } \sigma = W - \theta. \tag{1}$$

Where $\theta$ is the amount outstanding ("FlightSize") when the sending TCP function (e.g., `tcp_output()`) is called, $\sigma$ is the amount to send, starting from $nxt$. Such mode of operation has implications on the burstiness of the traffic produced by a TCP. These implications are well-known and recommendations in RFC6582 are made to limit the possible bursts of packets sent at once, for example, using an additional variable $maxburst$. In the present report, $W$, $\theta$ and $\sigma$ are in unit of packets (segments). Further, as $W_R$ never changes, we set $cwnd \leq W_R$ (*). No attempt is made to limit the data bursts.

TCP uses two methods to detect missing data: retransmission timeout and duplicate acks.

When the retransmission timer fires, a timeout occurs and $cwnd = 1$, $nxt = una$, thus $\sigma = 1$ and the missing packet is retransmitted. If this is the first retransmission of the packet, $S = \min(\lfloor \theta/2 \rfloor, 2)$. $S$ is left unchanged in case of multiple retransmissions of the same packet. The connection proceeds in slow-start mode. As a consequence, duplicate packets can be sent after a timeout; duplicates can also be lost, though.

The second method is via duplicate acks, typically three, which means that four consecutive acks request the same sequence number. From RFC5681, "*After receiving 3 duplicate ACKs, TCP performs a retransmission of what appears to be the missing segment, without waiting for the retransmission timer to expire.*" This is called fast retransmit. We abbreviate this event as TD, for Triple Duplicate acks. "*After the fast retransmit algorithm sends what appears to be the missing segment, the "fast recovery" algorithm governs the transmission of new data until a non-duplicate ACK arrives.*" With a small abuse, we abbreviate a successful fast recovery as FR. When a TD occurs, $S = \min(\lfloor \theta/2 \rfloor, 2)$ (same as for timeouts), and the missing packet is retransmitted. The previous value of $nxt$ is restored to allow the sending of new data with further arriving DAs and $cwnd = S + 3$. Note that no new data is sent here even if it were possible (specifically, when $\theta = 4$). For each subsequent duplicate ack received, $cwnd += 1$ and new data may be sent. Finally, "*When the next ACK arrives that acknowledges previously unacknowledged data, a TCP MUST set cwnd to ssthresh (the value set in step 2). This is termed "deflating" the window.*" At this point, depending on the TCP version used, FR occurs or not, more details are given below. Assume FR occurs; thus, $cwnd = S$. "Normal" operation resumes, that is, $cwnd$ is increased either to $W = S + 1/S$ if $W_{CA} = S$ or to $W = S + 1$ if $W_{CA} = S + 1$ and new data is sent depending on $\sigma = W - \theta$ being positive. In both cases, the connection proceeds in congestion avoidance mode.

If any one of the retransmissions is also lost then FR does not occur, instead, a timeout follows. We do not address this matter here.

For TCP Reno, FR occurs on receipt of the first ack requesting unacknowledged data. The loss of only one packet in the previous window of data has no further consequence. If there were $d$ drops, then another TD or a timeout follows. Under the assumption $W_{CA} = S + 1$, let $W$ be the window at which drops occur, let $\delta$ be the distance, in packets, between the first two drops, if $\delta < W - S - d$ then a timeout follows; otherwise, a TD follows if, in turn, enough acks return from the new packets sent during the TD-FR period.

For TCP NewReno (RFC6582), FR occurs when all data outstanding at the time TD occurred is acknowledged. Note that the loss of the new data sent between TD and FR lead to a new congestion event. Because NewReno is the focus of the present report, we give an example of operation in Figure 1.

---

* It is not advisable to do so when $W_R$ can change during the course of a connection.

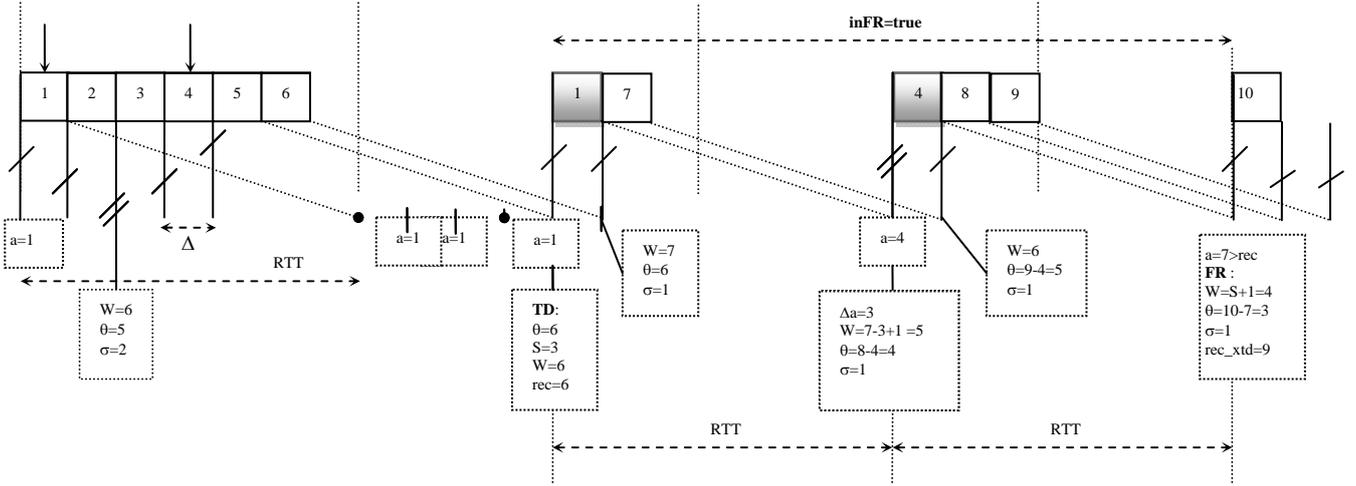

**Figure 1. Example of TD-FR with TCP NewReno from the sender perspective.**

Without loss of generality for the present purpose, we renumber the first loss as "1" and resynchronize time. When TD occurs (on the arrival of the ack from "5"), the missing packet is retransmitted, $recover = 6$, and both $S$ and $cwnd$ are adjusted. $W_R$ is large enough and does not change so that $W = cwnd$. The flag $inFR = true$. On receipt of the subsequent ack (from "6"), new data is sent. When the first loss is recovered, FR is not yet over for NewReno. Following RFC6582, the missing packet is retransmitted and a new data packet is sent. Further new data is sent on receipt of the ack from "7", which was new also. FR is over on the receipt of the acknowledgement $a = 7 > recover$; $inFR = false$, $cwnd = S + 1 < W_R$; normal operation resumes and new data is sent. In the same RTT, the receipt of the ack from "8" will lead to $cwnd += 1/cwnd$ as it must in congestion avoidance. For the purpose of collecting further statistical data on timeouts and TD, we record the highest sequence number sent during the TD-FR period in the variable $rec\_xtd$; if one or more of these new packets is lost a TD or a timeout will follow FR.

The example also leads to the following remarks.

(i)     On FR, the amount outstanding is $S$, if $W_{CA} = S$, no packet is sent. With $cwnd = S + 1/S$ on FR, it will catch up on receipt of the ack from "9" in the same RTT; this requires that neither "8" nor "9" are lost.

(ii)    The loss of any of the retransmissions will lead to a timeout in standard TCP.

(iii)   In congestion avoidance (with proper time resynchronization and a receiver acknowledging every packet), the window increase is one per RTT but the increase can occur almost any time in the RTT until the next packet drop. That remark is important for the analysis of section 5.

As concerns timeouts, RFC6582 states the following: "*After a retransmit timeout, record the highest sequence number transmitted in the variable recover, and exit the fast recovery procedure if applicable.*" Subsequent TD events are ignored if they do not cover more than $recover$. This is to avoid spurious TDs. Our TCP follows RFC6582 only for the first retransmission after a timeout. For subsequent retransmissions of the same packet $recover = 0$, thus enabling all TD events. Further, a $TDoff$ flag is used for the purpose of collecting timeout statistics.

We now give further useful details on the TD event until the receipt of all duplicate acks (DA) that elicited the TD. Assume that $\theta = W$ and that the number of losses in the window of size $W$ is $1 \leq d \leq W - 3$. After the TD event and the retransmission of the missing packet, the window increases with the arrival of DAs from $w = S + 3$ up to $w = \theta$, where no packets are sent ($w$ is used here to avoid confusion with $W$). Subsequent DAs allow the sending of new packets, the number of which is noted $N_{DA}$, provided the receiver window allows. Almost immediately we have the following results, whether queuing occurs or not.

$$W_{end} = W + S - d. \tag{2}$$

$W_{end}$ is the window when all "first" DAs have been received, and comes from $W_{end} = (S + 3) + (W - d) - 3$.

$$N_{DA} = \begin{cases} W_R - W & , W \leq W_R < W_{end} \\ S - d & , W_R \geq W_{end} \end{cases} \tag{3}$$

$N_{DA}$ is the number of new data packets sent with the receipt of 'first' DAs.

Although of secondary importance for our purpose, we also have $X = W - S - 3 \geq 0$, where $X$ is the number of DAs without sending (because $w \leq \theta$). Finally, $\theta_{end} = W + N_{DA} \leq W_R$, where $\theta_{end}$ is the amount outstanding when all "first" DAs have been received.

The number of packets sent during this period is $1 + N_{DA}$, including the retransmission.

The above results lead to the following comments.

(i)     Assume $W_R$ is large enough; for $d = 1$, the number of packets sent is $S \cong W/2$ in the RTT between TD and FR.

(ii)    If $W_R$, is not large enough, then, only $W_R - W$ new packets are sent. In particular, for a TD at $W = W_R$ a burst of $\lfloor W_R/2 \rfloor + 1$ packets is sent on FR.

(iii)   For $d > 1$, the number sent increases by one for each RTT corresponding to the retransmission and acknowledgement of a lost packet; provided no retransmission is lost again.

(iv)    As concerns $W_R$, the worst case is for $\Delta a = 1$ packet, which frees only one place in the receive buffer. Thus the sending of new packets may stop due to receive buffer limitation. As a consequence, a burst of packets will be sent on the subsequent FR.

(v)     The retransmission timer is set only at the time of TD; thus, up to $\lfloor RTO/RTT \rfloor$ previous losses can be retransmitted before a timeout occurs.

(vi)    With TDFR events only, the time of window increase in a RTT in normal operation depends on the previously sent $N_{DA}$ packets. Thus the sentence: "the window increases 'sometimes' in a RTT".

The final point of this section is the calculation of the retransmission timeout, which depends on the RTT of the connection, as defined in RFC6298. There are four points to consider: (i) Setting of the initial RTO (ii) Computing a proper RTO. (iii) RTT measurement. (iv) Setting the RTO for multiple retransmissions of the same packet.

(i). The initial RTO, when no RTT measurement has been made yet, is set to 1 second (it was previously 3 seconds). The "backing off" discussed below applies.

(ii). When RTT measurements, $R$, say, are available, the RTO is computed in the following two steps; assuming precise timing (zero clock granularity).

(a)     First measurement:
        $SRTT \leftarrow R, RTTVAR \leftarrow R/2, RTO \leftarrow \max(1\text{ s}, SRTT + 4 \times RTTVAR)$.

(b)     Subsequent measurements:
        $RTTVAR \leftarrow (1 - 1/4) \times RTTVAR + 1/4 \times |SRTT - R|$,
        $SRTT \leftarrow (1 - 1/8) \times SRTT + 1/8 \times R$,
        $RTO \leftarrow \max(1\text{ s}, SRTT + 4 \times RTTVAR)$.

With $R$ constant, the first measurement gives $RTO \leftarrow \max(1\text{ s},\ 3 \times R)$. On the tenth measurement, $RTO \leftarrow \max(1\text{ s},\ 1.113 \times R)$, and on the twentieth, $RTO \leftarrow \max(1\text{ s},\ 1.006 \times R)$, etc. Therefore, for a large connection, if $R > 1$ second $RTO = R$. A number of popular implementations use a $minRTO$ of 200 ms instead of 1 second for the setting of the RTO value.

(iii). RTT measurements are performed, at least, once per RTT outside of any loss event, and using new data. This means that retransmissions are not timed and measurements are not performed during TD-FR periods. On FR, normal operation resumes and the first segment sent is timed. Generally speaking, RTT measurements are taken once per "flight" at the beginning of the flight. This has no particular implication for the present work; however, spurious timeouts can occur with too large (in relative terms) $W_R$ and router buffers over slow links.

It is important not to confuse timed segments for the purpose of measuring the RTT and setting the retransmission timer on segments. Under normal operation, on retransmission after a timeout, and on FR, the timer must be set on each packet sent (timed or not) – in practice, when a burst of packets is sent, the timer is set on the first of the group. During a TD-FR period, the timer is set only on the first retransmission.

(iv). Backing off the RTO. When a packet is retransmitted the RTO is multiplied by two. When the same packet is retransmitted multiple times the RTO is multiplied again by two, up to four times ($2^5 = 64$) or up to the maximum RTO of 60 seconds. According to Windows implementation public documents, the retransmission of the same packet is limited to five times, at which point the connection is aborted. We do not consider this case here. Our full model (as other models) does not consider the maximum time, only the doubling up to five times.

Our TCP follows the above cited RFCs with the following specifics. The receiver acknowledges every packet received. The send buffer is "infinite". The initial window is 2 packets. The RTO value and RTT measurements are set with precise timing; that is, the $G$ of RFC6298 is zero. For TD events, the retransmit timer is set for the first retransmission only. Our TCP does not send new data on the first two duplicate acks before TD occurs (no "limited transmit"). We use $W_{CA} = S + 1$, which give better performance than the alternative. More precisely, the throughput is better by 4%, 14%, 15%, 13%, and 10% for $p = 1\%$, 5%, 10%, 15%, and 20%, respectively. Clearly, the choice of $W_{CA}$ is not without consequence, everything else being the same.

### 3.2 Network settings

The network setting used in this report is illustrated in Figure 2. There is a single long TCP connection using the links. Packets are dropped by the drop module on the receiver side. Acknowledgements are never lost. The router node has sufficient buffer so that no packets are dropped due to buffer overflow. A traffic module collects traffic information – arrivals and departures – at the entry and exit of the router for cross-checking purposes. The TCP sender and receiver are also instrumented to collect data for further checking and model validation. In order to collect data for the model developed here, each of the TCP packets includes additional information such as window, window change flag, type of retransmission; this information is put in the packet at the time it is sent. Obviously, such information is not included in a real TCP packet. In a simulator it is easily done for the purpose of modeling and evaluation.

With reference to Figure 2, we note $\Delta_1 = P/C_1$ and $\Delta = P/C$, the transmission times of data packets The round-trip time – not counting queuing – is $RTT = D + (P + a) \times (1/C_1 + 1/C)$, where $D$ is the – constant – round-trip propagation delay. RTT is the time required to send a packet and receive its acknowledgement. By linearity, RTT is the same from any point of view; sender output, node input, etc. In the present work, RTT is a constant and does not include possible queuing.

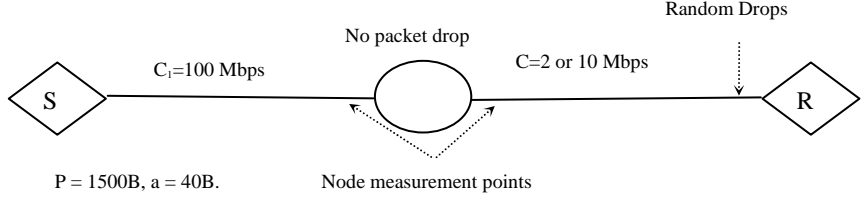

**Figure 2. Network setting.**

We let $\beta = RTT/\Delta$. $\lfloor \beta \rfloor$ gives the number of packets that can "fit" in the link of capacity $C$. $\beta$ is similar to the well-known $BDP = (\frac{D}{C})/P$.

We let $r = \frac{\lfloor \beta \rfloor}{W_R}$. $r \geq 1$, means that, in the long term, the link (path) is not saturated; $r < 1$ means that the link (path) is saturated.

When the link is saturated, the long-term queue content is $Q = W_R - \lfloor \beta \rfloor + 1$. The "1" accounts for the packet in transmission. The number of packets outstanding is $W_R$, $\lfloor \beta \rfloor$ are 'in the link', thus the difference is in the waiting queue and one in transmission. In order to be complete, we would have $RTT_Q = RTT + Q \times \Delta$, the RTT measured when queuing occurs. We do not use this observation here.

A second observation is that when the link is saturated, the number of packets in the link is $\lfloor \beta \rfloor$ per RTT. $\lfloor \beta \rfloor$ is also the number of returning acks per RTT. Therefore, for $W > \lfloor \beta \rfloor$, the window increase is less than one per RTT. For example, at $W = n\lfloor \beta \rfloor + x$, $x < \lfloor \beta \rfloor$, it requires $n \times RTT + x\Delta$ seconds to increase the window by one. Further note that each window increase adds one packet to the queue.

Table 1 gives the data for the different simulation scenarios we consider (with $\Delta_1 = P/C_1 = 120$ μs fixed).

**Table 1. Network parameters used in the report.**

| C (Mbps) | $\Delta$ (ms) | RTT (ms) | β (pkts) | $W_R$ (pkts) | r | Condition under $p = 0$ |
|---|---|---|---|---|---|---|
| 2 | 6 | 100 | 16.7 | 12 | 1.33 | Link not saturated |
| | | 100 | 16.7 | 24 | 0.67 | Link saturated, $Q = 9$ |
| | | 200 | 33.3 | 24 | 1.38 | Link not saturated |
| | | 200 | 33.3 | 44 | 0.75 | Link saturated, $Q = 12$ |
| 10 | 1.2 | 40 | 33.3 | 32 | 1.03 | Link not saturated |
| | | 50 | 41.7 | 44 | 0.93 | Link saturated, $Q = 4$ |
| | | 100 | 83.3 | 44 | 1.89 | Link not saturated |

The approaches and models developed below, as well as the various approximations, are compared with packet-level simulations according to the network parameters of Table 1. For each simulation, the connection transfers 10 millions packets when $p \geq 0.5\%$ and 20 millions packets otherwise (to increase the number of useful events). We verified that with such numbers there is negligible variability (within +/-0.2%) in the results of various simulations with the same parameters; we therefore compare our calculations with the results of one simulation only. Our implementation follows the details given in section 3.1. In particular, $S$ is calculated from the amount outstanding rather than the congestion window. Fortunately, although there are differences observed in our simulations, these differences are so small and rare that we can continue to use the congestion window as a substitute for the amount outstanding in models.

# 4. Simple send rate models

In this section we develop 'simple' send rate model that take into account only the TD-FR events. We further study their domains of applicability. Finally, we compare the theoretical results with simulations.

### 4.1 Square root law

The square root law [1], is widely used for obtaining the throughput of a long TCP connection in congestion avoidance.

In congestion avoidance and assuming a receiver window large enough, the sender window repeats cycles of increases between $W/2$ and $W$ and instantaneous decrease to $W/2$ when a drop occurs. Further, the window increases at the rate of one per RTT. The sender window gives the number of packets sent; the number sent in a cycle is thus $N = \frac{W}{2} + \left(\frac{W}{2} + 1\right) + \cdots + \left(\frac{W}{2} + \frac{W}{2}\right) = \frac{3}{4}W\left(\frac{W}{2} + 1\right)$. The duration of a cycle is $T = \left(\frac{W}{2} + 1\right)RTT$. The send rate is thus $SR = \frac{NP}{T} = \frac{3}{4}\frac{W \times P}{RTT}$.

With random i.i.d. packet drops and a drop probability $p$, the average number of packets appearing between drops is $1/p$. Relating $N$ and $1/p$ gives $N \cong \frac{3}{8}W^2 = 1/p$, which leads to $W = \sqrt{\frac{8/3}{p}} = W_{eq}$. $W_{eq}$ is the equilibrium window. Finally, $SR = \frac{P}{RTT}\sqrt{\frac{3/2}{p}} = \frac{P}{RTT}W_{eff}$, with $W_{eff} = \sqrt{\frac{3/2}{p}}$, the effective window.

Besides the congestion avoidance assumption, the above reasoning requires that the window increase is one per RTT (recall that RTT is a constant independent of queuing) and a receiver acknowledging every packet. This further requires that the connection path is not saturated; otherwise, only $\lfloor \beta \rfloor < W$ acks return per RTT and the increase is not one per RTT. With our notations, $r = \frac{\lfloor \beta \rfloor}{W_{eq}} \leq 1$ that is, $p \geq p_{min} = \frac{8/3}{\lfloor \beta \rfloor^2}$, which relates $p_{min}$ to the 'BDP' of the path.

We do not develop here the case of a saturated path.

In summary, for a receiver acknowledging every packet and no limitation on $W_R$, the square root law is:

$$SR_{sqrt} = \frac{P}{RTT}\sqrt{\frac{3/2}{p}}, \text{ with } W_{eff} = \sqrt{\frac{3/2}{p}}, \text{ for } p \geq p_{min} = \frac{8/3}{\lfloor \beta \rfloor^2}. \tag{4}$$

It is clear from the expressions above that the square root model is to be used with caution below $p_{min}$; the send rate grows without bounds when $p$ gets small. On the other hand, the domain of validity of the law extends quickly to small $p$ as $\beta^2$.

### 4.2 Linear law

The results presented in this section have been previously published in [21]; here, we develop a bit more on the matter.

Under the assumption that the receiver acknowledges every packet, that the sender window is limited to $W_R$, and that the path is not saturated, the linear law (in $p$) is:

$$SR_{lin} = \frac{P}{RTT} W_R \left[ 1 - \frac{W_R}{4} \left( 1 + \frac{W_R}{2} \right) p \right], \text{ with } W_{eff} = W_R \left[ 1 - \frac{W_R}{4} \left( 1 + \frac{W_R}{2} \right) p \right], \text{ for } p \leq p_{max} = \frac{2}{W_R^2}. \text{ (5)}$$

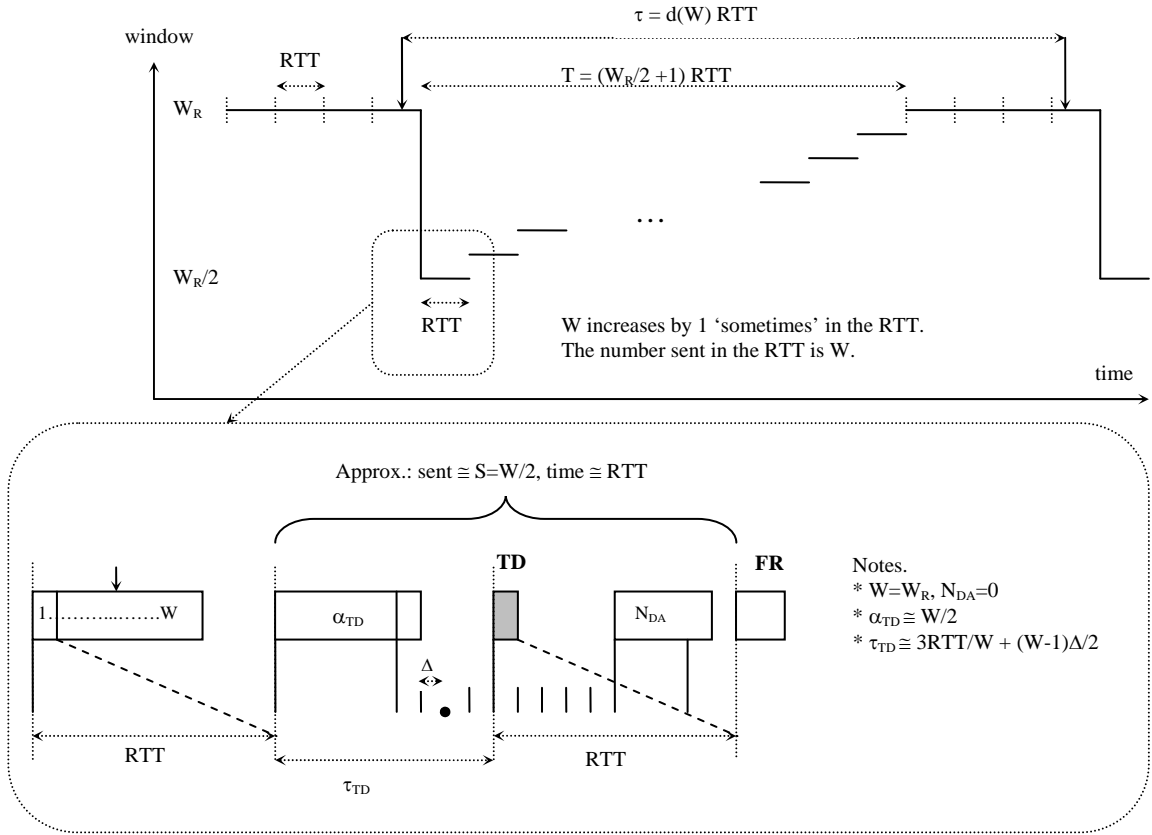

**Figure 3. Illustration of the approach for the linear law.**

The argument is illustrated in Figure 3.

To simplify notations, let $W = W_R$. Let $\tau = d(W) \times RTT$ be the average time between drops. Similarly to the case of the square root law, the number of packets sent in congestion avoidance until $W_R$ is reached for the first time is $N_{CA} = \frac{3}{4} W (\frac{W}{2} + 1)$ in the time $\gamma \times RTT$, with $\gamma = \left( \frac{W}{2} + 1 \right)$. After that, the number of packets sent is $[d(W) - \gamma] W$ in the time $(d(W) - \gamma) \times RTT$. The send rate is thus $SR = \frac{P}{RTT} \frac{N_{CA} + (d(W) - \gamma) W}{d(W)}$. It remains to calculate $d(W)$.

Now, $d(W)$ is the average number of RTTs between drops and $1/p$ is the average number of packets between drops. During the congestion avoidance phase the average number of packets sent is $\frac{3}{4} W$ per RTT, which we round up to $W$. With that approximation, the number of packets sent is $d(W) \times W$. Letting $1/p = d(W) \times W$ gives $d(W) = 1/pW$.

Plugging the expression for $d(W)$ into the $SR$ equation gives $SR = \frac{P}{RTT} \frac{\frac{3}{4} W (\frac{W}{2} + 1) + \left( 1/pW - \frac{W}{2} - 1 \right) W}{1/pW}$. The result above follows after a little algebra.

The result applies under the condition that $d(W) \geq \gamma$; otherwise the window does not return to $W_R$. The condition is rewritten as $\frac{1}{pW} \geq \frac{W}{2} + 1$, which is approximated as $p \leq \frac{2}{W^2}$.

Replacing back $W$ by $W_R$ in both the send rate expression and the condition gives the result given at the beginning.

For completeness, when the path is saturated (recall, no buffer overflow only random drops), we have the following linear law (here, time is not counted in units of RTT but in units of $\Delta = P/C$):

$$SR = C(1 - \text{Kp}) \text{ and } W_{\text{eff}} = \beta(1 - \text{Kp}) \text{ for } p \leq p_{\max}, \tag{6}$$

with somewhat complicated expressions for K and $p_{\max}$.

Under the further conditions $W_R - \lfloor\beta\rfloor > 3$ and $\lfloor W_R/2\rfloor < \lfloor\beta\rfloor$,

$$\text{K} = (\lfloor\beta\rfloor - \lfloor W_R/2\rfloor + 2)\beta - \frac{(\lfloor\beta\rfloor - \lfloor\frac{W_R}{2}\rfloor + 1)(\lfloor\beta\rfloor + \lfloor W_R/2\rfloor + 2)}{2} - 1, \tag{7}$$

and

$$1/p_{\max} = (\lfloor\beta\rfloor - \lfloor W_R/2\rfloor + 2)\beta + W_R + \frac{(W_R - \lfloor\beta\rfloor + 1)(W_R + \lfloor\beta\rfloor)}{2} - 1. \tag{8}$$

The above conditions mean that the link does not go idle before the retransmission of the missing packet and that the link does not become saturated on FR; that is, $W_{FR} = S + 1 = \lfloor W_R/2\rfloor + 1$, and that many packets are sent back to back because none are outstanding.

### 4.3 Domains of validity

For convenience, we summarize the previous results. Under the conditions that the receiver acknowledges every packet and that the path (link) is not saturated, we have:

$$SR_{sqrt} = \frac{P}{RTT}\sqrt{\frac{3/2}{p}}, \text{ with } W_{\text{eff}} = \sqrt{\frac{3/2}{p}}, \text{ for } p \geq p_{\min} = \frac{8/3}{\lfloor\beta\rfloor^2}.$$

$$SR_{lin} = \frac{P}{RTT}W_R\left[1 - \frac{W_R}{4}\left(1 + \frac{W_R}{2}\right)p\right], \text{ with } W_{\text{eff}} = W_R\left[1 - \frac{W_R}{4}\left(1 + \frac{W_R}{2}\right)p\right], \text{ for } p \leq p_{\max} = \frac{2}{W_R^2}.$$

Remarks:

- The maximum long term send rate is $C$, which gives $W_{\text{effmax}} = C \times \frac{RTT}{P} = \beta$.
- If we let $W_R = \beta$, that is, at the onset of path saturation, we have that $p_{\max} = \frac{2}{\beta^2} < p_{\min} = \frac{2.67}{\beta^2}$.
- A path not saturated means that $W_R \geq \lfloor\beta\rfloor$. Therefore, $p_{\max} < p_{\min}$ for $1 \leq r \leq \sqrt{4/3} = 1.15$.

The following Table 2 gives the network parameters used for the validation, the conditions of applicability of the simple models, and the settings for which timeout statistics are discussed below.

The comparison between the $W_{\text{eff}}$ calculated and measured from simulations is illustrated in Figure 4.

From Figure 4 [(c) is a zoom of (a)], it can be seen that both the square root and the linear models are accurate in their respective domain of validity and up to $p \cong 1 - 2\%$. The linear model is accurate up to $p_{\max}$. The case (2MR200W44), where $\beta = 33.3$, is shown in Figure 4 (b) as an illustration that when the path can be saturated, the linear model without saturation does not apply; rather the more complicated linear model under saturation must be used.

**Table 2. Network setting for the evaluation of the simple TCP throughput models.**

| C (Mbps) | Δ (ms) | RTT (ms) | β (pkts) (= W$_{eff}$ max) | W$_R$ (pkts) | p$_{min}$(%) sqrt | p$_{max}$(%) linear | Comments |
|---|---|---|---|---|---|---|---|
| 2 | 6 | 100 | 16.7 | 12 | 1.04 | 1.39 | No sqrt. TO stats. |
| | | 100 | 16.7 | 24 | 1.04 | 0.35 (not sat) 0.64 (sat) | sqrt after linear. |
| | | 200 | 33.3 | 24 | 0.24 | 0.35 | sqrt after linear. TO stats. |
| | | 200 | 33.3 | 44 | 0.24 | 0.10 (not sat) 0.20 (sat) | sqrt after linear. |
| 10 | 1.2 | 40 | 33.3 | 32 | 0.24 | 0.20 | sqrt after linear. TO stats |
| | | 50 | 41.7 | 44 | 0.16 | 0.10 (not sat) | sqrt after linear. TO stats |
| | | 100 | 83.3 | 44 | 0.039 | 0.10 | sqrt after linear. TO stats |

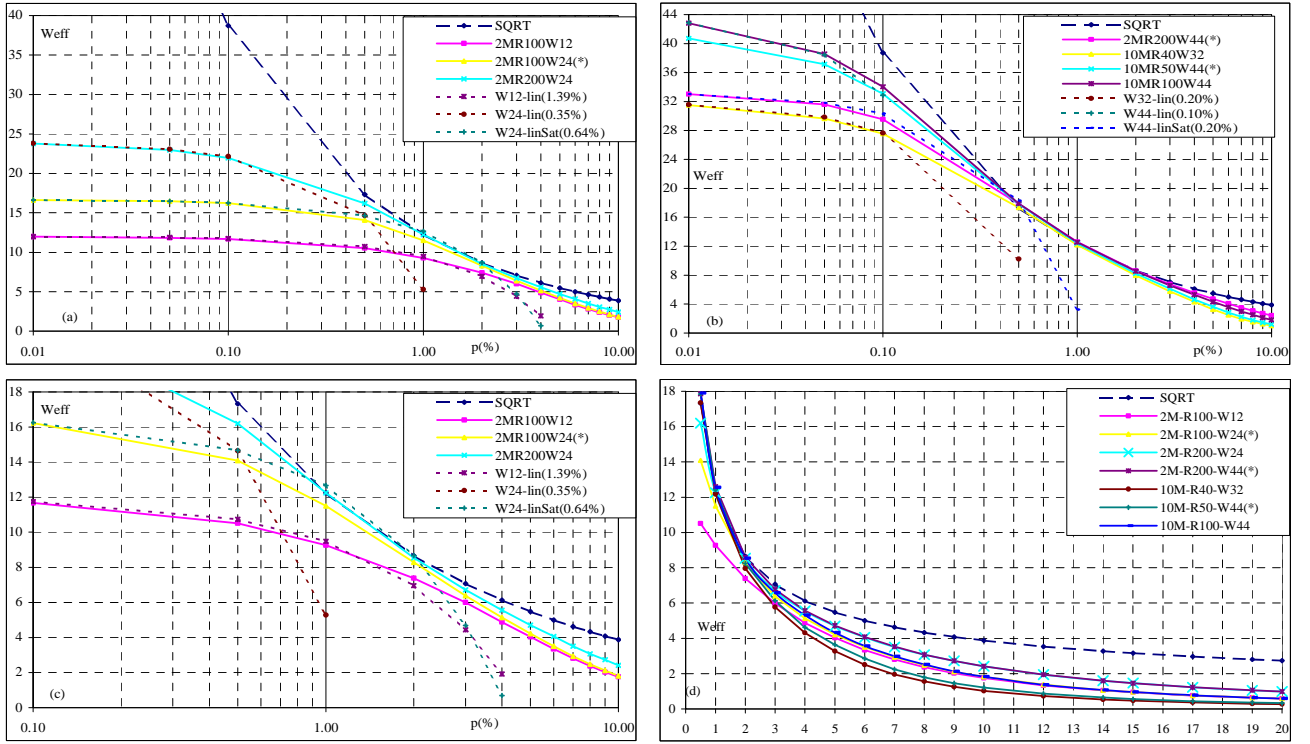

**Figure 4. Comparison of simulations and calculations for the simple models.**

We make the following observation. From Figure 4 (b) and (d), it can be seen that $W_{eff} \cong 12$ for $p = 1\%$, and $W_{eff} \cong 8$ for $p = 2\%$, whatever large is $W_R$. On the other hand, from Figure 4 (c), it can be seen that the linear model is accurate up to about $p = 2\%$. The error is slightly above 2% at $p = 0.5\%$ ($W_{eff} = 10.7$) and $p = 1\%$ ($W_{eff} = 9.5$), and $-6\%$ at $p = 2\%$, ($W_{eff} = 7$). This suggests that using parallel connections with a $W_R = 12$ each to transfer a large file can increase the throughput as well as increase the link utilization for $p \cong 0.5 - 2\%$. Under random packet drops, connections will see the same drop probability; thus, each connection will have the same $W_{eff}$, which in turn, add up to link capacity.

The above discussion lead to consider three regimes for the TCP send rate under random packet drops: linear regime, "sqrt" regime, and timeout regime, depending on $p$. This is illustrated by the following Figure 5.

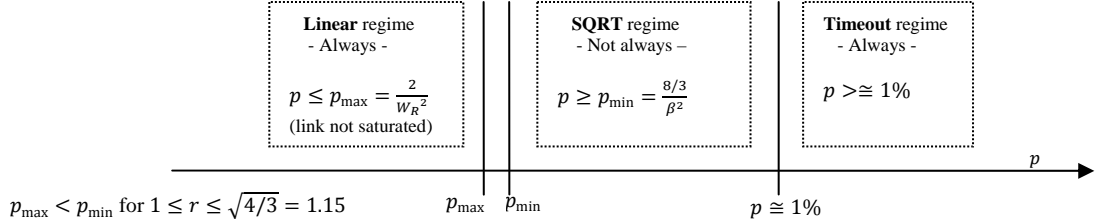

**Figure 5. Domain of validity for the different TCP throughput models.**

The "sqrt" regime is not always present as exemplified by the case for $W_R = 12$. On the other hand, the linear regime may not be "visible" for large $W_R$ and $\beta$. In this case, the "sqrt" regime applies for a wide range of $p$ up to the timeout regime.

We split the analysis of simulation results in two parts: $p$ below 1% and above.

In Figure 6 we give the average simulation results (using NewReno) for four network settings and for $p$ up to 1%. "LE" stands for 'loss events' as measured at the sender, the labels "%rxtFR(n)" indicate the percentage of retransmitted packets on successful FRs. We make the following comments. Loss events are essentially made of TDs. For $p$ up to 0.1 %, there is essentially one packet dropped and retransmitted. Starting from $p = 0.5\%$, double packet drops cannot be ignored. At $p = 1\%$ and above, timeouts cannot be ignored although they constitute only about 1% of the loss events; this is because $minRTO$ is 1 second. This gives us a first indication that the timeout regime may start at $p \cong 1\%$.

| p | %TD/drps | %LE/drps | %TD/LE | %FR/TD | %rtxFR(1) | %rtxFR(2) | %rtxFR(3) | %TO/LE |
|---|---|---|---|---|---|---|---|---|
| 0.01 | 99.950 | 99.950 | 100.000 | 100.000 | 99.950 | 0.200 | | 0.000 |
| 0.05 | 98.647 | 98.713 | 99.935 | 99.934 | 98.713 | 1.268 | 0.020 | 0.065 |
| 0.1 | 98.044 | 97.620 | 99.890 | 99.888 | 97.615 | 2.333 | 0.080 | 0.110 |
| 0.5 | 91.549 | 92.058 | 99.448 | 99.454 | 91.740 | 7.885 | 0.363 | 0.553 |
| 1 | 87.375 | 88.453 | 98.795 | 98.840 | 87.978 | 11.165 | 0.815 | 1.218 |

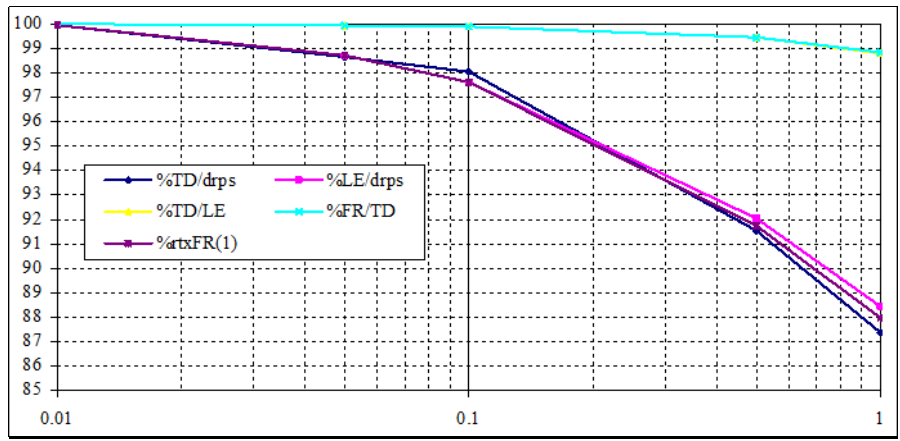

**Figure 6. Simulation results for $p$ up to 1%.**

Turning to the second part of the analysis, in Figure 7, we give more details for $p$ ranging from 1% to 10%. The results represent the averages for the five settings mentioned in Table 2 as 'TO stats'.

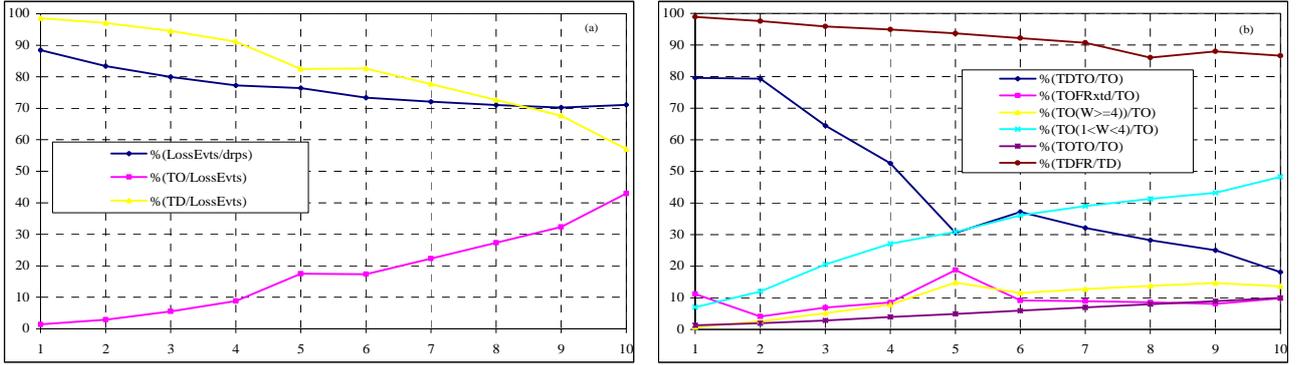

**Figure 7. Timeouts analysis for $p$ between 1 and 10 percent .**

From Figure 7 (a) we see that the loss events (TD + TO) rate greatly underestimates the packet drop rate. As $p$ increases, the importance of TDs decreases (approximately linearly) from 98.6% down to less than 70% of the loss events while the importance of timeouts increase (approximately as $p^{1.5}$) from 1.4% up to 43%. At $p = 2\%$, the proportion of TOs is 3%, at $p = 3\%$ that proportion is 5.5%. Thus, we set at three percent the threshold [TO/(TD + TO)] for the timeout regime, which, in turn, correspond to $p <\cong 2\%$.

Figure 7 (b) shows that the proportion of TDs followed by FRs is high, slightly smaller than $1 - p$. The label 'TOFRxtd' indicates timeout occurring after a successful FR due to the loss of new packets sent in the TD-FR period. Not counting the unexplained peak at $p = 5\%$, the contribution of this type of TO is up to 10%. Below $p \cong 5\%$, the major contribution to timeouts comes from TDTO, that is, the loss of retransmissions in TD-FR periods. Above $p \cong 5\%$, the major contribution to timeouts comes from timeouts at low window, $W = 2$ or $W = 3$, that is, the loss of one packet leads to a timeout.

To conclude this section, we investigated the case where retransmissions are never dropped, which is possible with a simulator. The results are shown in Figure 8.

Figure 8 (a) suggest that the possible improvements of the send rate are not clearly visible. Thus, timeouts due to low window still govern the performance of TCP under high packet drop rates. Figure 8 (b), gives the percent improvement when either the retransmissions in TD-FR periods (flag rtxTD) or all types of retransmissions (flag rtxALL, meaning rtxTD and rtxTO) are not lost. When TD-FR retransmissions are not lost, the improvement is only up to 12%. While it seems possible to develop algorithms to detect and retransmit lost retransmissions during TD-FR events (e.g., using the SACK version or by analyzing the return of acks in NewReno), it remains to decide whether the improvement is worth the effort. On the other hand, the detection of lost retransmissions after timeout is impossible as only one packet is sent and lost. The time spent without sending anything is 2, 4, 8, etc., seconds. TCP implementations that use a $minRTO$ of 200 ms instead of 1 s certainly fare better than those following strictly the standards, although, with a constant RTT of 200 ms, spurious timeouts must occur frequently in TD-FR. Another possibility is to use "limited transmit". Yet another possibility is to use an adaptive $minRTO$ depending on the current window instead of the fixed 1 second value. It is not in the scope of the present work to develop further that matter.

We now turn to the full model.

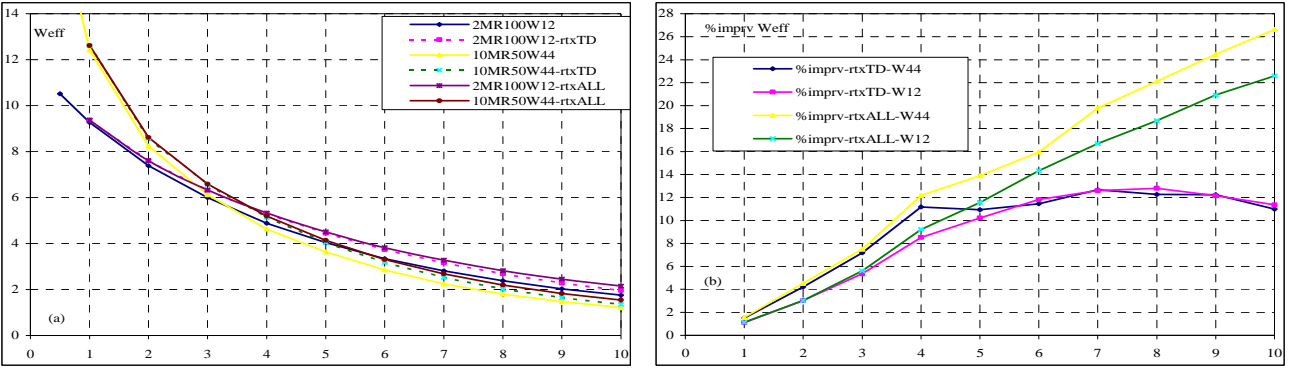

**Figure 8. W$_{eff}$ when retransmissions are not dropped.**

# 5. Full send rate model

In this section, we first give the exact equations governing the send rate of a TCP connection. We then give approximations for the NewReno TCP variant under the assumption that the path is not saturated.

### 5.1 Exact model

The TCP throughput or send rate is defined by the following equation [2]:

$$SR = \lim_{t \to \infty} \frac{N_t}{t}, \text{ where } N_t \text{ is the amount sent until time } t.$$

Let $\frac{N_t}{t} = \frac{N_{RTT} + N_{TD} + N_{TDFR} + N_{TDTO} + N_{TO}}{T_{RTT} + T_{TD} + T_{TDFR} + T_{TDTO} + T_{TO}}$.

Note that this decomposition is exhaustive. We note $N_x$ the total number of packets sent and $T_x$ the total time spent up to time $t$ in each of the 5 basic elements (see Figure 9 below): RTT, TD, TDFR, TDTO, and TO. RTT means "normal operation" in a RTT. There may be none (under saturation) or more than one window change in the RTT – in slow-start there may be several window changes in the same RTT, we use the last one. TD means Triple Duplicate acknowledgement. TDFR means TD followed by a successful fast recovery. TDTO means TD followed by a timeout due to the loss of a retransmission. TO means "direct timeout".

Let us index the totals above by the sender window, $W$, and let $N = \sum_{W=1}^{W=W_R} N(W)$. $N(W)$ is the number of times the sender window changes to $W$, not counting the changes between TD (excluded) and FR (included) events. $N(W_R)$ gives the number of changes to $W_R$ either from $W_R - 1$ or after receiving $W_R$ acks. $W = 1$ means that a timeout occurred; therefore, $N(1)$ gives the number of timeouts. In the limit $t \to \infty$, $P(1) = N(1)/N$ gives the probability of timeout. The indexing by $W$ is also exhaustive.

To simplify the discussion, we consider the case where the path is not saturated and give indications relative to the case where it is saturated.

Write $N_{RTT} = \sum_W n_{RTT}(W) = N \sum_W \frac{N(W)}{N} \frac{n_{RTT}(W)}{N(W)}$, which, in the limit, will give (with an abuse of notation)

$$N_{RTT} = \sum_W P(W)\alpha_{RTT}(W). \tag{9}$$

NOTE: When the link is saturated, that is, $W > \lfloor \beta \rfloor$, $n_{RTT}(W) = \lfloor \beta \rfloor$, independent of $W$ (recall that $RTT$ is a constant). We may relax the assumption in a probabilistic manner after the calculation of $P(W)$, i.e.,

with $P(W > \lfloor\beta\rfloor) \leq \varepsilon$ as a function of $p$, so that the possible saturation has little impact on the results of the model; see also section 6.

Similarly, $T_{RTT} = \sum_W t_{RTT}(W) = N \sum_W \frac{N(W)}{N} \frac{t_{RTT}(W)}{N(W)}$. Whether the link is saturated or not, we have, by definition, $t_{RTT}(W) = RTT$, a constant independent of $W$; in the limit (with an abuse of notation),

$T_{RTT} = \sum_W P(W)\tau_{RTT}(W)$, that is,

$$T_{RTT} = RTT. \tag{10}$$

For TD events,

$N_{TD} = \sum_W n_{TD}(W) = N \sum_W \frac{N(W)}{N} \frac{N_{TD}(W)}{N(W)} \frac{n_{TD}(W)}{N_{TD}(W)}$, which will give (with an abuse of notation)

$$N_{TD} = \sum_W P(W)P_{TD}(W)\alpha_{TD}(W). \tag{11}$$

$$T_{TD} = \sum_W P(W)P_{TD}(W)\tau_{TD}(W). \tag{12}$$

NOTE: With our network settings, the quantities that depend on the path being saturated or not are $\alpha_{TD}(W)$ and $\tau_{TD}(W)$.

For TDFR events,

$N_{TDFR} = \sum_W n_{TDFR}(W) = N \sum_W \frac{N(W)}{N} \frac{N_{TD}(W)}{N(W)} \frac{N_{TDFR}(W)}{N_{TD}(W)} \frac{n_{TDFR}(W)}{N_{TDFR}(W)}$; in the limit (with an abuse of notation),

$$N_{TDFR} = \sum_W P(W)P_{TD}(W)P_{TDFR}(W)\alpha_{TDFR}(W). \tag{13}$$

$$T_{TDFR} = \sum_W P(W)P_{TD}(W)P_{TDFR}(W)\tau_{TDFR}(W). \tag{14}$$

For TDTO events,

$$N_{TDTO} = \sum_W P(W)P_{TD}(W)P_{TDTO}(W)\alpha_{TDTO}(W). \tag{15}$$

$$T_{TDTO} = \sum_W P(W)P_{TD}(W)P_{TDTO}(W)\tau_{TDTO}(W). \tag{16}$$

Finally, for timeouts, TO events,

$$N_{TO} = \sum_W P(W)P_{TO}(W)\alpha_{TO}(W). \tag{17}$$

$$T_{TO} = P(1)P_{TO}(1)\tau_{TO}(1) + \sum_{W=2} P(W)P_{TO}(W)\tau_{TO}(W).$$

Letting $\tau_{TOTO} = P_{TO}(1)\tau_{TO}(1)$ gives

$$T_{TO} = P(1)\tau_{TOTO} + \sum_{W=2} P(W)P_{TO}(W)\tau_{TO}(W). \tag{18}$$

The approach is illustrated in Figure 9.

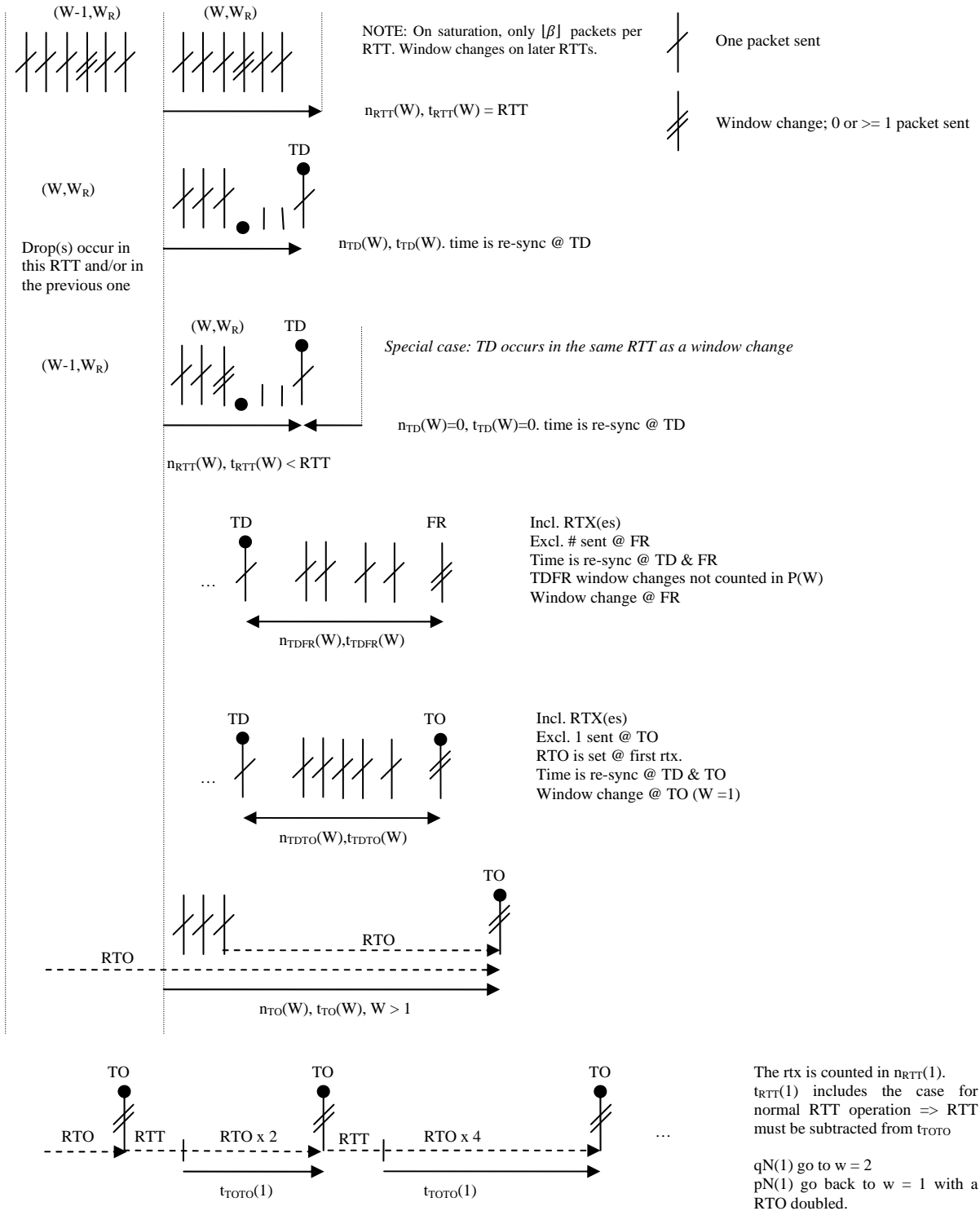

**Figure 9. Illustration of the basic elements of the model.**

It remains to calculate the 16 quantities,

$P(W),$

$\alpha_{RTT}(W), \tau_{RTT}(W),$

$P_{TD}(W), \alpha_{TD}(W), \tau_{TD}(W),$

$P_{TDFR}(W), \alpha_{TDFR}(W), \tau_{TDFR}(W),$

$P_{TDTO}(W), \alpha_{TDTO}(W), \tau_{TDTO}(W),$

$P_{TO}(W), \alpha_{TO}(W), \tau_{TO}(W), \tau_{TOTO}.$

The calculation of these quantities depends on the TCP version considered.

The version considered here is TCP NewReno with $W_{CA} = S + 1$, meaning that slow start continues until $W \leq S$. In particular, on FR, $W = S + 1$.

### 5.2 Calculation of P(W)

We assume pure random i.i.d. packet drops meaning that an arriving packet is dropped with probability $p$ and is not dropped with probability $q = 1 - p$, independently of previous events or other parameters. Nothing happens if no packet appears. Note also that drops are only random, not due to buffer overflow.

We start by writing $P(W_R) = q^{W_R} P(W_R) + q^{W_R-1} P(W_R - 1)$, which means that none of the $W_R$ packets sent at $W_R$ are dropped, that is, $W_R$ acks are received. Another possibility is that no drop occur at $W_R - 1$, i.e., $W_R - 1$ acks are received thus increasing the window to $W_R$.

Next,

$$P(W) = q^{W-1} \times P(W-1)$$
$$+ I_{2W-1 \leq W_R} \times P_{FR}(2W-1) \times P(2W-1)$$
$$+ I_{2W-2 \leq W_R} \times P_{FR}(2W-2) \times P(2W-2).$$

$I_x$ is the indicator function of the condition $x$. There are 3 ways to reach $W$: From $W-1$ and no drops. From a TD at $2(W-1)$, or at $2(W-1)+1$ in which cases $S = W-1$ and on the following FR, $W_{FR} = S + 1 = W$.

$P(2) = qP(1)$. That is, the retransmission after a timeout is not lost.

$P(1)$ is given by $P(1) = 1 - P(2) - \cdots - P(W_R)$.

After this first step, we write, $P(W) = A_{W,W-1} \times P(W-1)$.

For $W = W_R$, $A_{W,W-1} = \frac{q^{W_R-1}}{1-q^{W_R}}$.

For $W$ such that $2W - 2 > W_R$, $A_{W,W-1} = q^{W-1}$.

For $W$ such that $2W - 2 = W_R$, $A_{W,W-1} = \frac{q^{W-1}}{1-P_{FR}(2W-2) \times A_{2W-2,W}}$, with $A_{2W-2,W} = \prod_{i=W}^{2W-2} A_{i+1,i}$.

For $W$ such that $2W - 1 \leq W_R$, $A_{W,W-1} = \frac{q^{W-1}}{1-P_{FR}(2W-2) \times A_{2W-2,W} - P_{FR}(2W-1) \times A_{2W-1,W}}$.

For $W = 2$, $A_{2,1} = q$.

With the above, we have $P(1) = \frac{1}{1+A_{2,1}+A_{3,1}+\cdots A_{W_R,1}}$.

We use $P_{FR}(W) = q^W(Wp + \frac{W(W-1)}{2}p^2)$, meaning that we consider only one or two drops in a sequence of $W$ packets, and none or the retransmission is lost. Under random i.i.d. packet drops we found this approximation sufficient.

Finally, $P(W) = A_{W,1}P(1) = P(1) \times \prod_{i=2}^{W} A_{i,i-1}$.

NOTES:

- The calculation of $P(W)$ is for a TCP version such that $S = \left\lfloor \frac{W}{2} \right\rfloor$ on TD and $W_{FR} = S + 1$.
- For a version such that $W_{FR} = S + 1/S$ (still a window change on FR) one must replace $2W - 2$ by $2W$ and $2W - 1$ by $2W + 1$.
- $P(W)$ does not depend on whether the link is saturated or not.
- $P(W)$ does not depend on timing considerations such as the RTT.
- $P(W)$ does not depend on the packet size (but $p$ may).

Example with $W_R = 8$:

$P(8) = \frac{q^7}{1-q^8}P(7) = A_{8,7}P(7); P(7) = q^6 P(6) = A_{7,6}P(6); P(6) = q^5 P(5) = A_{6,5}P(5);$

$P(5) = \frac{q^4}{1-P_{FR}(8)A_{8,7}A_{7,6}A_{6,5}}P(4) = A_{5,4}P(4); P(4) = \frac{q^3}{1-P_{FR}(7)A_{7,6}A_{6,5}A_{5,4}-P_{FR}(6)A_{6,5}A_{5,4}}P(3) = A_{4,3}P(3);$

$P(3) = \frac{q^2}{1-P_{FR}(5)A_{5,4}A_{4,3}-P_{FR}(4)A_{4,3}}P(2) = A_{3,2}P(2); P(2) = qP(1) = A_{2,1}P(1); $ finally,

$P(1) = \frac{1}{1+A_{2,1}+A_{3,1}+\cdots A_{8,1}}.$

The following Figure 10, Figure 11, and Figure 12 show the comparison between calculations and simulations.

For $p = 1\%$, calculations and simulations are almost indiscernible for $W_R = 12, 24, 32, 44$. Note the modes at the respective receiver window size.

For $p = 5\%$, calculations and simulations are almost indiscernible for $W_R = 12$. However, differences in the tail (below about 1% and $W > 12$) begin to appear at $W_R = 24$ and are more marked for $W_R = 32$ and $W_R = 44$. These differences remain unexplained; they are not due to using the congestion window (in calculations) as a surrogate for the amount outstanding (in simulations). Note that for a given $p$ the calculated distributions are the same up to the common $W$.

For $p = 10\%$, the difference between calculations and simulation are still small for $W_R = 24$. Differences are more marked for $W_R = 32$ and $W_R = 44$, below about 1% and $W > 10$.

Fortunately, for larger $p$, this discrepancy and its effects are reduced due to the fact that the sender window does not grow high.

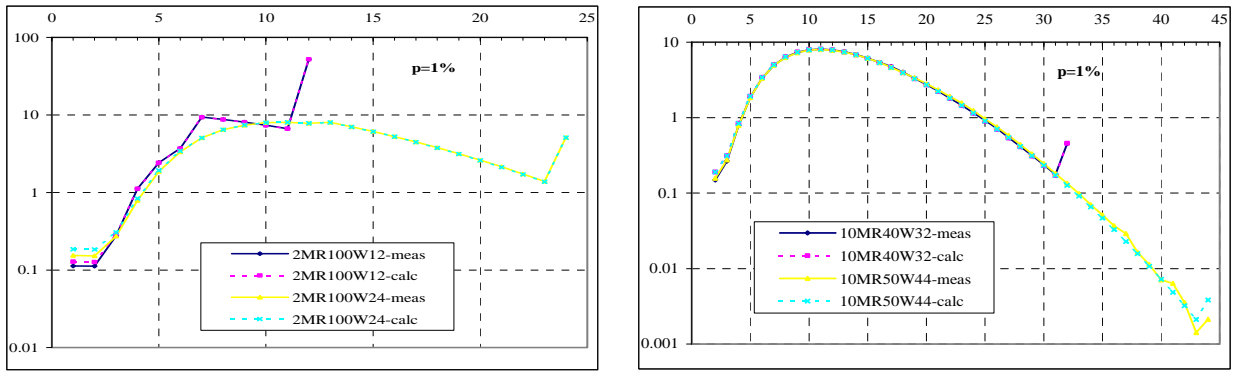

**Figure 10. Comparison between calculated and measured $P(W)$ for $p = 1\%$.**

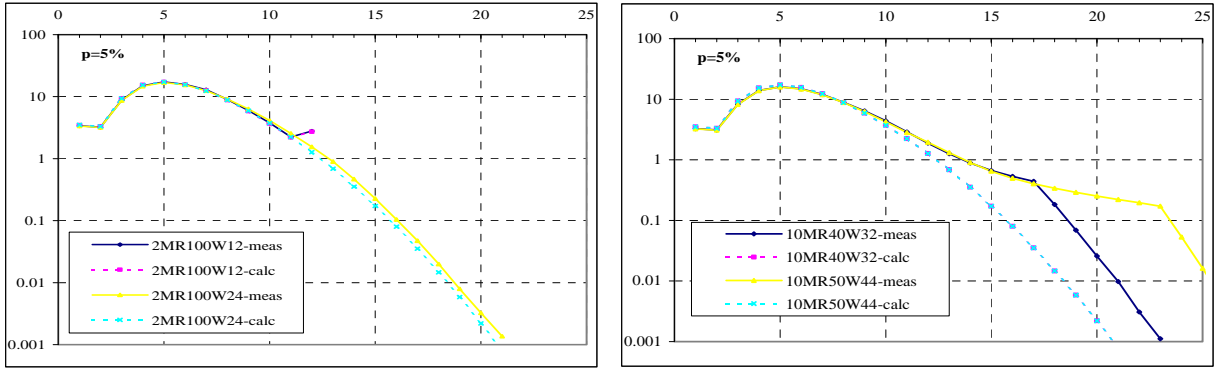

**Figure 11. Comparison between calculated and measured $P(W)$ for $p = 5\%$.**

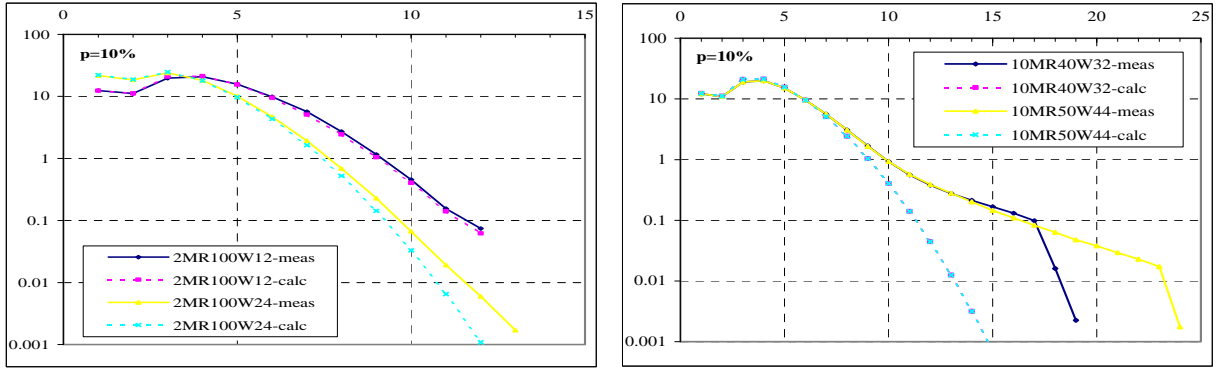

**Figure 12. Comparison between calculated and measured $P(W)$ for $p = 10\%$.**

NOTE: In the following sections, we explicitly assume that the link is not saturated at any $W$.

### 5.3 Calculation of $\alpha_{RTT}(W)$ and $\tau_{RTT}(W)$

We have already calculated $\tau_{RTT}$ above as

$$\tau_{RTT} = RTT. \tag{19}$$

This expression does not take into account the possibility of a TD occurring in the same RTT as a window change (see Figure 9).

For $\alpha_{RTT}$ we use the following approximation,

$$\alpha_{RTT} = <W> = \sum_{i=1}^{W_R} iP(i). \tag{20}$$

This is an approximation because TCP can send less than $W$ packets in a RTT; see the example in the next section. It can also send more; see the example in section 5.6, which also shows that a TD can occur at $W = 3$. The more and the less do not compensate.

NOTE: All approximations below assume that a TD can occur only with $W \geq 4$.

### 5.4 Calculation of $P_{TD}(W)$, $\alpha_{TD}(W)$, and $\tau_{TD}(W)$

A TD occurs whenever between 1 and $W - 3$ packets are dropped in a series of $W$; these $W$ packets are not necessarily in the same RTT, see Figure 13.

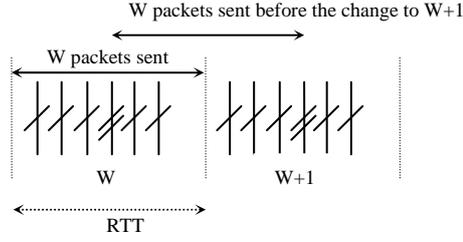

**Figure 13. W packets may appear in contiguous RTTs.**

For i.i.d. packet drops, we use the following approximation, which gives the probability of 1 or 2 drops among $W$ packets.

$$P_{TD}(W) = Wpq^{W-1} + \frac{W(W-1)}{2}p^2q^{W-2}. \tag{21}$$

For $\alpha_{TD}(W)$ and $\tau_{TD}(W)$ we use the following approximations:

$$\alpha_{TD}(W) = \frac{W-1}{4} \text{ and } \tau_{TD}(W) = \frac{2RTT}{W} + \frac{W-1}{2}\Delta, \text{ with } \Delta = P/C. \tag{22}$$

First, there are $W$ equally likely cases for the loss of one packet in a window of $W$ in the same RTT. The total for the number of packets sent in the next RTT and until TD occurs is $0 + 1 + \cdots + (W-1) = W(W-1)/2$. The total for the time elapsed until TD is $3RTT + (0 + 1 + \cdots + (W-1))\Delta = 3RTT + W(W-1)\Delta/2$. The average number sent is $(W-1)/2$ and the average time is $3RTT/W + (W-1)\Delta/2$.

However, there are other cases to consider. For example, consider a loss at $W = 6$ or at $W = 7$. On TD, $S = 3$ for both cases. Let $W_R \geq 9$. With duplicate acknowledgements, 2 new packets, $p1$ and $p2$, say, are sent before FR occurs. On FR, $W = 4$, the amount outstanding is 2 packets and 2 new packets are sent, $p3$ and $p4$, say. If neither $p1$ nor $p2$ are dropped, 2 more packets will follow in the same RTT, $p5$ and $p6$, say. We thus fall back to the case above if a drop occurs on $p3$, or $p4$, or $p5$, or $p6$. In this case, 4 packets are sent in the RTT; the same as $W$. However, if $p1$ is dropped, a new TD will occur triggered by the receipt of the acknowledgement of $p4$, the time to count is $\Delta$. Note that only 2 packets were sent in the RTT at $W = 4$. A similar reasoning applies to the loss of $p2$; we note also that only 3 packets are sent in the RTT at $W = 4$. Returning to our approximation, we see that there are $4 + 2 \times 2 = 2W$ cases to consider. These cases are clearly not equally likely; letting them being so is also part of the approximation. We have not attempted to perform an exact analysis at all windows. In particular, there are not always $2W$ cases.

### 5.5 Calculation of $P_{TDFR}(W)$, $\alpha_{TDFR}(W)$, and $\tau_{TDFR}(W)$

After a TD occurred, it is necessary that all retransmissions are not lost in order for a FR to occur. New data packets sent between TD and FR may be dropped, though. We calculated $P_{FR}(W)$ in section 4.1, but here we know that a TD occurred.

$$P_{TDFR}(W) = 1 - P_{TDTO}(W), \tag{23}$$

with $P_{TDTO}(W)$ given in the next section.

For $\alpha_{TD}(W)$ we use the following expressions

$$\alpha_{TDFR}(W) = 1 + N_{DA}(W) \text{ with } N_{DA}(W) = \begin{cases} W_R - W & , W \leq W_R < W_{end} = W + S - 1 \\ S - 1 & , W_R \geq W_{end} \end{cases}. \quad (24)$$

Note that this expression accounts for one drop only. For NewReno and two drops, the expression is somewhat more complicated and does not add much to the precision of the global result. Note also that the leading "1" accounts for the retransmission.

As concerns $\tau_{TD}(W)$, we use

$$\tau_{TD}(W) = RTT(1 + I_{W>4} \times \frac{W-1}{2} \frac{p}{q}). \quad (25)$$

In this case, we take into account the possibility of two drops for increased precision and little more complication. Note also that $\frac{W-1}{2} \frac{p}{q} = \frac{P(2 \text{ drops})}{P(1 \text{ drop})}$.

### 5.6 Calculation of $P_{TDTO}(W)$, $\alpha_{TDTO}(W)$, and $\tau_{TDTO}(W)$

For $P_{TDTO}(W)$ we use the following

$$P_{TDTO}(W) = p \left( 1 + I_{W>4} \times \frac{W-1}{2} \frac{p}{q} \right). \quad (26)$$

The 'average' probability of TDTO is larger than $p$ when there is more than one drop to recover.

For $\alpha_{TDTO}(W)$ we use the following

$$\alpha_{TDTO}(W) = \min \left( 1/p, W_R - S - 2, N_{DA} \times \lceil RTO/RTT \rceil \right). \quad (27)$$

If $N_{DA} = 0$, which occurs if the TD occurs at $W = 4$ or $W = W_R$, no packets are sent between TD and TO.

If $N_{DA} > 0$, new packets are sent until the retransmission timer fires or the amount outstanding reaches $W_R$ or because packets are dropped and no ack returns to the sender.

For large $W_R$ ($W_R = 32$ and $44$) the above approximation gives a result that is too small compared with simulations. However, as discussed in section 6, the influence of the number sent during this event is of secondary importance.

We only consider the TCP version that sets the retransmission timer on the first retransmission (on TD) therefore,

$$\tau_{TDTO}(W) = RTO. \quad (28)$$

NOTE: We have experimented with a version that sets the timer on each retransmission and observed only very minor influence (within +/-0.2%, the same as simulation variability) on the global results.

We now give an example where after a TDTO TCP can send 4 packets at $W = 3$ or 5 packets at $W = 4$. Suppose that after a TD packet $p0$ is lost again, thus a timeout occurs. Suppose that packets $p1$, $p2$, $p3$, $p4$, and $p5$ were also sent with packets $p1$, $p2$, and $p4$ dropped. On TO $p0$ is retransmitted and $S = 2$; if $p6$ was also previously sent, $a = p1$, $W = 2$ and $p1$, $p2$ are sent again. In the third round, the ack from $p1$ is $a = p2$, $W = 3$, $\theta = (3 - 2) = 1$, $\sigma = W - \theta = 2$, and $p3$, $p4$ are sent. The ack from $p2$ is $a = p4$, $W = 3 + 1/3$, $\theta = (5 - 4) = 1$, thus 2 more packets are sent, making a total of 4 at $W = 3$. With $S \geq 3$, the total would be 5 at $W = 4$. Note that acks are always increasing here. Further, if one of the 4 packets sent is dropped, we would have a TD at $W = 3$ with $\theta = 4$.

### 5.7 Calculation of $P_{TO}(W)$, $\alpha_{TO}(W)$, and $\tau_{TO}(W)$

In this section we only consider the cases $W = 2$ up to $W = 5$ for which 1, 2, or 3 drops lead to a TO.

For $W = 2$,

$$P_{TO}(2) = p, \alpha_{TO}(2) = 0, \text{ and } \tau_{TO}(2) = RTO - RTT. \tag{29}$$

If the first packet of the group of 2 is dropped – whatever happens to the second packet – a TO occurs. For $\tau_{TO}$, note that the timer was set when sending this first packet; the first RTT is counted as normal operation.

The expressions are more complicated for $W = 3$. To first order in $p$, i.e., with only one drop.

$$P_{TO}(3) = \left[\frac{P(2)}{P(3)}qp\right] + [p + qp] + \left[\frac{P(4)}{P(3)}P_{FR}(4)(1 + q + q^2)p\right] + \left[\frac{P(5)}{P(3)}P_{FR}(5)(1 + q)p\right]. \tag{30}$$

$W = 3$ can be reached from $W = 2$, where two packets are sent. The first packet must not be dropped (to reach $W = 3$) but the second must. Only up to two acks are received thus a TO follows. Now, assume the two packets sent at $W = 2$ are not dropped. If $S > 2$, then $W = 4$ on receipt of the second ack and we are not concerned here. If $S = 2$, two packets are sent at $W = 3$, on receipt of the second ack (from $W = 2$), $W = 3 + 1/3$. To have a TO event, either the first packet must be dropped or, if it passes, the second must be dropped. If none are dropped, $W = (3 + 1/3) + 1/3 + 1/3 = 4$; again, we are not concerned.

$W = 3$ can be also be reached from a TDFR at $W = 4$. On FR, $W = (2 + 1) = 3$ and three packets are sent because none are outstanding. To have a TO, either the first, or the second, or the third must be dropped.

Finally, $W = 3$ can be also be reached from a TDFR at $W = 5$. In this case $N_{DA} = 1$. Two packets are sent on FR. Again, there are only two possibilities to consider for this case.

This makes a total of 8 cases, not equally likely. See below the approximations for $\alpha_{TO}(3)$, and $\tau_{TO}(3)$.

To have a TO at $W = 4$, requires, at least, two packet drops. We use the following approximation.

$$P_{TO}(4) = 2 \times \binom{4}{2}p^2q^2 = 2 \times 6p^2q^2. \tag{31}$$

To have a TO at $W = 5$, requires, at least, three packet drops. We use the following approximation:

$$P_{TO}(5) = 3 \times \binom{5}{3}p^3q^2 = 3 \times 10p^3q^2. \tag{32}$$

For $W > 2$, we use the following approximations for $\alpha_{TO}(W)$, and $\tau_{TO}(W)$.

$$\alpha_{TO}(W) = \frac{W-1}{4} \text{ and } \tau_{TO}(W) = RTO - 0.5RTT. \tag{33}$$

### 5.8 Calculation of $\tau_{TOTO}$

With reference to Figure 9 bottom, with probability $pq$ the time spent in timeout is $2RTO - RTT$. With probability $p^2q$ the time spent is $(2 + 4)RTO - RTT$. With probability $p^3q$ the time spent is $(6 + 8)RTO - RTT$. With probability $p^4q$ the time spent is $30RTO - RTT$. With probability $p^5q$ the time spent is $62RTO - RTT$, which corresponds to the fifth doubling. The doubling stops now and the time spent remains at this value for any subsequent timeout on the same packet.

The average time is given by

$$\tau_{TOTO} = 2RTO \times pq \times (1 + 3p + 7p^2 + 15p^3 + 31p^4 \sum_{i=0}^{\infty} p^i) - RTT \times pq \sum_{i=0}^{\infty} p^i.$$

The sum is elementary and we have

$$\tau_{TOTO} = 2RTO \times pq \times (1 + 3p + 7p^2 + 15p^3 + \frac{31p^4}{q}) - pRTT$$

Finally, the proportion of time spent in multiple retransmissions of the same packet can be rewritten as

$$P(1)\tau_{TOTO} = P(1) \times p \times [2RTO \times (1 + 2p + 4p^2 + 8p^3 + 16p^4) - RTT].$$

However, simulations indicate that a better approximation is given by:

$$P(1)\tau_{TOTO} = q \times P(1) \times p \times [2RTO \times (1 + 2p + 4p^2 + 8p^3 + 16p^4) - RTT]. \tag{34}$$

Notes: (i) The packet retransmitted multiple times is included in $< W >$. (ii) The RTT has been accounted for in $\tau_{RTT}$ given above; therefore, it must be removed from the time spent in retransmitting the same packet. (iii) The above reasoning is valid for an implementation that limits the number of doubling to five independently of the time spent.

The model is summarized in Table 3.

**Table 3. Summary of exact approach and approximations for TCP send rate.**

| Exact approach | Approximations for NewReno TCP with $S = \left\lfloor \frac{W}{2} \right\rfloor$ on TD and $W_{FR} = S + 1$. |
|---|---|
| $SR = \frac{N_{RTT} + N_{TD} + N_{TDFR} + N_{TDTO} + N_{TO}}{T_{RTT} + T_{TD} + T_{TDFR} + T_{TDTO} + T_{TO}}$ | Notes: - TD can occur only when $W \geq 4$. <br> - $W \leq \lfloor \beta \rfloor$, path not saturated. |
| | $P(W)$ is numerically calculated as per section 5.2. |
| $N_{RTT} = \sum_W P(W)\alpha_{RTT}(W)$ <br><br> $T_{RTT} = \sum_W P(W)\tau_{RTT}(W)$ | $\alpha_{RTT}(W) = <W>$ <br><br> $\tau_{RTT}(W) = RTT$ |
| $N_{TD} = \sum_W P(W)P_{TD}(W)\alpha_{TD}(W)$ <br><br> $T_{TD} = \sum_W P(W)P_{TD}(W)\tau_{TD}(W)$ | $P_{TD}(W) = Wpq^{W-1} + \frac{W(W-1)}{2}p^2q^{W-2}$ <br><br> $\alpha_{TD}(W) = \frac{W-1}{4}$ <br><br> $\tau_{TD}(W) = \frac{2RTT}{W} + \frac{W-1}{2}\Delta$ |
| $N_{TDFR} = \sum_W P(W)P_{TD}(W)P_{TDFR}(W)\alpha_{TDFR}(W)$ <br><br> $T_{TDFR} = \sum_W P(W)P_{TD}(W)P_{TDFR}(W)\tau_{TDFR}(W)$ | $P_{TDFR}(W) = 1 - P_{TDTO}(W)$ <br><br> $\alpha_{TDFR}(W) = 1 + \begin{cases} W_R - W, & W \leq W_R < W_{end} = W + S - 1 \\ S - 1, & W_R \geq W_{end} \end{cases}$ <br><br> $\tau_{TD}(W) = RTT(1 + I_{W>4}\frac{W-1}{2}\frac{p}{q})$ |
| $N_{TDTO} = \sum_W P(W)P_{TD}(W)P_{TDTO}(W)\alpha_{TDTO}(W)$ <br><br> $T_{TDTO} = \sum_W P(W)P_{TD}(W)P_{TDTO}(W)\tau_{TDTO}(W)$ | $P_{TDTO}(W) = p(1 + I_{W>4}\frac{W-1}{2}\frac{p}{q})$ <br><br> $\alpha_{TDTO}(W) = \min(1/p, W_R - S - 2, N_{DA} \times \lceil RTO/RTT \rceil$ <br><br> $\tau_{TDTO}(W) = RTO$ |
| $T_{TO} = P(1)\tau_{TOTO} + \sum_{W=2} P(W)P_{TO}(W)\tau_{TO}(W)$ | $P_{TO}(2) = p, \alpha_{TO}(2) = 0, \tau_{TO}(2) = RTO - RTT.$ <br><br> $P_{TO}(3) = \left[\frac{P(2)}{P(3)}qp\right] + [p + qp] + \left[\frac{P(4)}{P(3)}P_{FR}(4)(1 + q + q^2)p\right] +$ <br> $\left[\frac{P(5)}{P(3)}P_{FR}(5)(1 + q)p\right]$ <br><br> $P_{TO}(4) = 2 \times \binom{4}{2}p^2q^2 = 2 \times 6p^2q^2$ <br><br> $P_{TO}(5) = 3 \times \binom{5}{3}p^3q^2 = 3 \times 10p^3q^2$ <br><br> $\alpha_{TO}(W) = \frac{W-1}{4}; \tau_{TO}(W) = RTO - 0.5RTT, W > 2$ <br><br> $P(1)\tau_{TOTO} = q \times P(1) \times p \times$ <br> $[2RTO \times (1 + 2p + 4p^2 + 8p^3 + 16p^4) - RTT]$ |

## 6. Validation of the full model

The simulations performed confirm that both the number of packets sent and the time spent in sending them follow (a) the decomposition in the five basic elements of the exact model. (b) The further indexing with the sender window of the exact model.

In order to cross-check results, the send rate is measured in four different manners: via the traffic analysis module, via the sender TCP, via the number measured in the five basic elements, and via the measurements through the sender window indexing. All four ways give the same results.

The following Figure 14 shows the percent error between simulations and the model as a function of the packet drop probability from 0.5% to 20%. (The label xM means C=xMbps, Ry means RTT=y ms, and Wz means $W_R$=z packets).

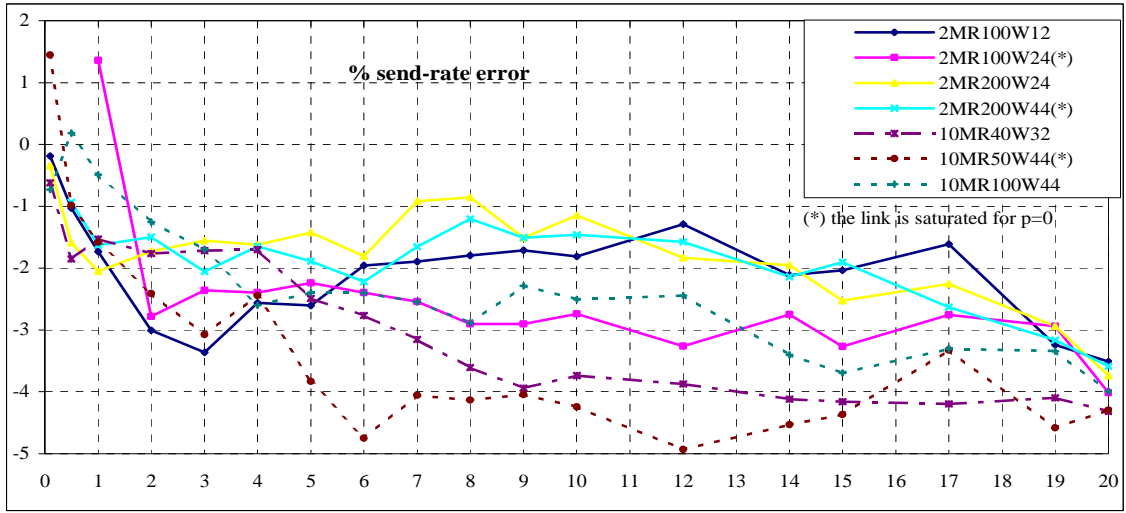

**Figure 14. Comparison between simulations and calculations of the TCP throughput.**

For the cases where the link is never saturated, the error is between 0 and $-5\%$. For the three cases where the link can be saturated, the model still does a good job for $p$ small but large enough. For the settings (10M, R50, W44), for $p$=0.1%, we calculate $P(W > 41) = 40.8\%$, while for $p = 0.5\%$, $P(W > 41) = 0.9\%$, which represent a drastic drop, thus the link is hardly saturated. For the settings (2M, R100, W24), for $p$=1%, we calculate $P(W > 16) = 24.2\%$; the model works nicely. We leave for future work the determination of the threshold $p_{th}$, above which the model can be applicable even when the link can be saturated at lower $p$.

We now turn to a more detailed analysis of the results along the line of the basic elements. Table 4 below gives the details for the worst case (10M, R50, W44) for $p = 5, 10, 15$, and 20%.

For each of the basic elements, the table gives the number of packets sent and the sum, the associated time, in ms, and the sum. The first row gives the results of calculations. The second row the measured values. The third row gives the percent proportion of the measured values. The fourth row gives the percent error of the sums between calculation and measurements. The fifth row gives the weighted percent error between calculations and measurements. The send rate (SR) is obtained by multiplying the number sent by the packet size (1500B).

Several remarks are in order.

- The model gives results that are always larger than measurements. The error is up to 6% for the numbers sent and up to 10% for the time spent. The send rate being a ratio the errors subtract. Yet the model is still reliable as indicated by the actual values and the weighted error row.
- As concerns the number sent, the major contribution comes from the number sent per RTT in "normal" operation.
- As concerns the time spent, the relative importance of timeouts, either direct or by lost retransmission, grows as $p$ increases, from a total of 19% at $p = 5\%$ to 46% and 35%, respectively, at $p = 20\%$. Note that the $minRTO$ is set to 1 second here. Some modern implementations set this value at 200 ms.
- The importance of the TDTO events, either in the number sent or the time spent, although not negligible, is minimal with respect to the other timeout events. Thus the exact modeling of this event is of secondary importance.

**Table 4. Detailed comparison of model and simulations.**

| | #inRTT | s-TOdir | s-TD | s-TDTO | s-TDFR | sum | t-Norm | rtx-lost | t-TOdir | tTD | t-TDTO | t-TDFR | sum | SR(kbps) |
|---|---|---|---|---|---|---|---|---|---|---|---|---|---|---|
| | | | | | | | 10M-R50-W44 -p5 | | | | | | | |
| calc | 5.854 | 0.010 | 0.403 | 0.274 | 0.797 | 7.338 | 50.000 | 3.221 | 17.719 | 4.926 | 14.952 | 14.204 | 105.023 | 838.476 |
| meas | 5.386 | 0.008 | 0.293 | 0.340 | 0.897 | 6.924 | 45.498 | 3.344 | 16.092 | 3.000 | 14.333 | 13.420 | 95.686 | 868.273 |
| %totMeas | 77.789 | 0.111 | 4.230 | 4.912 | 12.958 | | 47.549 | 3.495 | 16.817 | 3.136 | 14.979 | 14.025 | | |
| %err | | | | | | 5.991 | | | | | | | 9.757 | -3.432 |
| %wErr | 6.769 | 0.028 | 1.598 | -0.954 | -1.450 | | 4.705 | -0.129 | 1.701 | 2.013 | 0.647 | 0.820 | | |
| | #inRTT | s-TOdir | s-TD | s-TDTO | s-TDFR | sum | t-Norm | rtx-lost | t-TOdir | tTD | t-TDTO | t-TDFR | sum | SR(kbps) |
| | | | | | | | 10M-R50-W44 -p10 | | | | | | | |
| calc | 3.883 | 0.046 | 0.289 | 0.247 | 0.508 | 4.973 | 50.000 | 27.278 | 87.469 | 5.380 | 30.246 | 13.611 | 213.984 | 278.863 |
| meas | 3.615 | 0.040 | 0.227 | 0.346 | 0.553 | 4.781 | 47.060 | 26.834 | 79.568 | 3.731 | 27.804 | 11.840 | 196.835 | 291.449 |
| %totMeas | 75.628 | 0.833 | 4.743 | 7.227 | 11.569 | | 23.908 | 13.633 | 40.423 | 1.895 | 14.126 | 6.015 | | |
| %err | | | | | | 4.018 | | | | | | | 8.712 | -4.318 |
| %wErr | 5.594 | 0.139 | 1.294 | -2.068 | -0.941 | | 1.494 | 0.226 | 4.014 | 0.838 | 1.240 | 0.900 | | |
| | #inRTT | s-TOdir | s-TD | s-TDTO | s-TDFR | sum | t-Norm | rtx-lost | t-TOdir | tTD | t-TDTO | t-TDFR | sum | SR(kbps) |
| | | | | | | | 10M-R50-W44-p15 | | | | | | | |
| calc | 2.977 | 0.083 | 0.187 | 0.175 | 0.286 | 3.707 | 50.000 | 78.988 | 164.071 | 4.210 | 34.729 | 9.840 | 341.839 | 130.141 |
| meas | 2.852 | 0.071 | 0.148 | 0.204 | 0.295 | 3.570 | 48.702 | 76.740 | 149.632 | 3.031 | 29.305 | 7.812 | 315.223 | 135.912 |
| %totMeas | 79.882 | 1.993 | 4.146 | 5.708 | 8.271 | | 15.450 | 24.345 | 47.469 | 0.962 | 9.297 | 2.478 | | |
| %err | | | | | | 3.839 | | | | | | | 8.444 | -4.246 |
| %wErr | 3.497 | 0.324 | 1.080 | -0.801 | -0.261 | | 0.412 | 0.713 | 4.581 | 0.374 | 1.721 | 0.643 | | |
| | #inRTT | s-TOdir | s-TD | s-TDTO | s-TDFR | sum | t-Norm | rtx-lost | t-TOdir | tTD | t-TDTO | t-TDFR | sum | SR(kbps) |
| | | | | | | | 10M-R50-W44-p20 | | | | | | | |
| calc | 2.465 | 0.104 | 0.114 | 0.114 | 0.150 | 2.947 | 50.000 | 158.948 | 221.631 | 2.872 | 31.452 | 6.290 | 471.193 | 75.055 |
| meas | 2.420 | 0.087 | 0.086 | 0.108 | 0.148 | 2.848 | 49.475 | 152.617 | 202.904 | 2.023 | 24.456 | 4.516 | 435.989 | 78.399 |
| %totMeas | 84.973 | 3.042 | 3.002 | 3.786 | 5.196 | | 11.348 | 35.005 | 46.539 | 0.464 | 5.609 | 1.036 | | |
| %err | | | | | | 3.464 | | | | | | | 8.074 | -4.266 |
| %wErr | 1.576 | 0.617 | 1.001 | 0.208 | 0.062 | | 0.120 | 1.452 | 4.295 | 0.195 | 1.605 | 0.407 | | |

## 7. Conclusion

In the present work, we have analyzed and modeled the send rate of long connections that use standard TCP.

Our goal was to obtain models for the performance that can be obtained with TCP under the main assumptions of random packet drops (with probability $p$), and limited receiver window ($W_R$).

We have delimited three regimes depending on $p$ and $W_R$, as illustrated in Figure 5. Following common knowledge, the timeout regime starts at $p > \cong 1\%$. We experimentally determined that this threshold corresponds to timeouts constituting about 3% of the loss events. In the timeout regime, models used for congestion avoidance do not apply (for a compliant TCP). For $p < \cong 1\%$, where congestion avoidance dominates the performance of TCP, we have revisited the square root law and shown that it is applicable for $p$ above a threshold which depends on the BDP. Further, when taking into account the finiteness of the receiver window, we have shown that a linear law in $p$ applies up to a $p_{max}$, which we also calculate for both cases of an unsaturated and saturated path. When the linear and square root regimes overlap, the linear regime is to be applied. As shown in this report, simple expressions for the send rate and conditions of applicability allow the easy calculation of TCP performance in congestion avoidance.

In section 5, we have developed an exact approach for the calculation of TCP send rate usable in any regime. This exact approach requires the calculation of 16 quantities, which depend on the TCP version used. Fortunately, some of these quantities can be simple to obtain.

We have then provided and discussed approximations for the send rate of TCP NewReno. A by-product of these calculations is the distribution of the sender window, which turns out to be independent of any timing (RTT) and saturation considerations. Unfortunately, the full model does not yield analytic expressions although it does not require sophisticated numerical methods. Its use is therefore reserved to the timeout regime and off-line use.

A detailed comparison and discussion of the full model results for NewReno with simulation results has been provided in section 6.

We leave for future work the detailed analysis for the case where saturation occurs. We also leave for the future the application to the case of non standard TCP implementations.